\documentclass[%
 aip,
 jap,%
 amsmath,amssymb,
 reprint,%
]{revtex4-1}

\usepackage{graphicx}
\usepackage{dcolumn}
\usepackage{bm}
\usepackage{braket,color}
\newcommand{\Argum}[1]{\ensuremath{\! \left( #1 \right)}}

\newcommand{\mc}[1]{\ensuremath{\mathcal{ #1 }}}  
\newcommand{\TextSub}[1]{\ensuremath{_{\mbox{\tiny{#1}}} } }  
\newcommand{\TextSup}[1]{\ensuremath{^{\mbox{\tiny{#1}}} } }  
\newcommand{\Tr}{\mbox{Tr}}
\usepackage[caption=false]{subfig}

\begin{document}

\preprint{AIP/123-QED}

\title{Non-Markovian Models of Environmentally-driven Disentanglement in Molecular Charge Qubits}


\author{Shengyang Zhou}

\author{Enrique P. Blair}%
 \email{Electronic Address: Enrique\_Blair@baylor.edu}
\affiliation{ 
Affiliation: Electrical and Computer Engineering Department, Baylor University, Waco, TX, USA, 76798
}%

\date{\today}

\begin{abstract}
Models of quantum disentanglement are developed for nanometer-scale
molecular charge qubits (MCQs). Two MCQs, $A$ and $B$, are prepared
in a Bell state and separated for negligible $A$-$B$ interactions.
Interactions between the local environment and each MCQ unravels $A$-$B$
entanglement during coherent system+environment evolution. Three models
are used for dynamics: (1) a previously-developed, numerical model,
in which both $AB$ and environment $\mc{E}$ are modeled explicitly;
(2) an exact, semi-analytic model, in which only the dynamics of $AB$
are calculated, and (3) an approximate model developed from the semi-analytic
model and assumptions about randomness in $\mc{E}$. In the approximate
model, the non-zero coherences of the density operator for $AB$ decay
with a Gaussian time dependence. This provides a time scale for system
dynamics in the exact models as well. This time scale is related directly
to the strength of AB-$\mc{E}$ interaction. This time scale describes
cases where environmental interaction with one target MCQ is dominant,
generalizing a previous time scale applicable only when both MCQs
have roughly the same strength of interaction with the local environment.
Entanglement is measured using two-qubit correlation functions, the
dynamics of which are used to demonstrate the effectiveness of the
time scale. The early-time decay of coherences and the loss of entanglement
is well-characterized as Gaussian, a behavior that Markovian models
for memoryless environments cannot capture. The approximate Gaussian
model may be used to describe the dynamics of MCQ disentanglement
under the influence of environments modeled here, as well as other
environments where randomness is present.
\end{abstract}

\maketitle

\section{\label{sec:level1}Introduction}

Quantum computing promises new ways to process information and to
efficiently solve problems that are difficult or impossible for classical
computers.\citep{FeynmanQuantumSimulation1982,FeynmanGenesis1985}
Such applications include Shor's algorithm \citep{shor1994algorithms}
for defeating a widely-used encryption scheme, Grover's search algorithm,\citep{GroverSearchUnpub}
simulating quantum systems,\citep{FeynmanQuantumSimulation1982} and
optimization problems.\citep{adiabaticQC} Quantum cryptography promises
provably secure methods for sharing information.\citep{BB84,QuantumCryptography1991}
Entanglement between qubits is an essential resource in both quantum
computation and communication, but it is easily unraveled by qubit-environment
interactions.\citep{jozsa2003role}

Several physical implementations exist for quantum bits (qubits),
and still others could be invented. This paper focuses on molecular
charge qubits (MCQs), which could be implemented using $\pi$-cojugated
block copolymers\citep{2013_MCQ} or multi-metal-centered mixed-valence
molecules, suitable also for a general-purpose classical computing
paradigm known as quantum-dot cellular automata (QCA).\citep{LentScience2000,QCA_QuantComp_2001,QCA_at_molecular_scale}
Quality factors of $\sim10^{3}-10^{4}$ have been reported for MCQ
systems,\citep{2013_MCQ} making it feasible to process information
using MCQs.

In this paper, the dynamics of disentanglement are studied in MCQs
using computational and analytic methods. Here, a double-quantum-dot
(DQD) molecule provides an MCQ. A remotely-separated target pair of
MCQs is prepared in a Bell state for maximal entanglement. Vast spatial
separation eliminates Coulomb coupling between the target MCQs. Each
MCQ in the pair is allowed to interact Coulombically with its local
environment, which consists of $M$ charge-neutral DQD molecules.
This is the starting point for a time evolution, over which entanglement
in $AB$ is quantified using quantum correlation functions. Here,
the time dependence of disentanglement is found, along with a characteristic
time scale.

This work generalizes a previously-found time scale for environmentally-driven
disentanglement in the target Bell pair.\citep{blair2018entanglement}
Previous work was constrained to a regime in which the strength of
local environmental interactions was approximately equal for each
of the two target qubits. The previously-used time scale does not
generalize to cases where one MCQ in the target pair suffers the dominant
environmental interaction. In this paper, a more general time scale
found.

A previously-developed numerical model\citep{blair2018entanglement}
for the dynamics of disentanglement in $AB$ is reviewed, and an exact,
semi-analytic model is developed in Section \ref{sec:model}. Additionally,
the semi-analytic model is used with assumptions about randomness
in $\mc{E}$ to obtain an approximate model for the dynamics of disentanglement,
as well as to obtain a time scale characteristic of those dynamics.
The time scale is related directly to energies of interaction between
each target MCQ and its local environment and also characterizes the
dynamics of the exact models. Quantum correlation functions are used
to quantify entanglement in the target MCQ pair and to demonstrate
the effetiveness of the new time scale in characterizing the dynamics
of disentanglement. The dynamics of disentanglement are seen to have
a Gaussian form unattainable using Markovian models of a memoryless
environment. The approximate Gaussian model for disentanglement could
be used to describe $AB$ dynamics not only in the environments studied
here, but also in other randomly-arranged non-Markovian environments.

\section{Models of Disentanglement\label{sec:model}}

\subsection{A Molecular Charge Qubit\label{subsect:MCQ}}

A mixed-valence compound such as diferrocenyl acetylne (DFA) can function
as a molecular DQD.\citep{2013_zwitterionicQCA_DQD,molecularQCAelectronTransfer}
Here, two iron centers provide redox centers, each of which functions
as a molecular quantum dot. While the DFA molecule must be singly-ionized
to provide useful charge states for this application, other charge-neutral
(zwitterionic) molecules are under study for both molecular charge
qubits and for energy-efficient, beyond-CMOS classical computing applications.\citep{2011_zwitterions,2013_zwitterionicQCA_DQD,Christie15}
In this paper, charge-neutral DQD molecules similar to DFA are considered.

\begin{figure}
\begin{centering}
\includegraphics{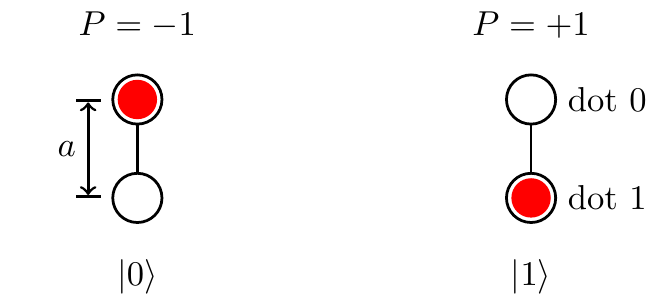}
\par\end{centering}
\caption{Localized electronic states of a molecular double quantum dot (DQD)
system provide the two classical states of a qubit. Black circles
represent the two quantum dots, and a connecting bar indicates a tunneling
path. A red disc represents the mobile electron.\label{fig:BasisStates}}
\end{figure}

Two charge-localized states of a molecular DQD provide the computational
basis states for a single MCQ (See Figure \ref{fig:BasisStates}).
Here, one mobile electron occupies one of two quantum dots. Also,
a fixed charge $+e/2$ (not pictured) is assumed to reside at each
dot, providing net charge neutrality for each DQD. Here, $e$ is the
fundamental charge, and the dots are treated as charged points separated
by distance $a$.
It will be helpful to quantify the charge state of a DQD in a single
number, the polarization, $P$, given by $P=\braket{\hat{\sigma}_{z}}$,
where $\hat{\sigma}_{z}$ is one of the Pauli operators $\{\hat{\sigma}_{x},\hat{\sigma}_{y},\hat{\sigma}_{z}\}$.

\subsection{A Bell Pair}

The system of interest, $AB$, is a target pair of entangled molecular
charge qubits, designated $A$ and $B$. The pair $AB$ is prepared
in a Bell state as the initial state of the time evolution:
\begin{equation}
\ket{\Psi_{AB}\left(0\right)}=\frac{1}{\sqrt{2}}\left(\ket{0_{A}}\otimes\ket{0_{B}}+\ket{1_{A}}\otimes\ket{1_{B}}\right)\label{eq:BellState}
\end{equation}
Henceforth, a more compact notation is used: $\ket{\Psi_{AB}\left(0\right)}=\left(1/\sqrt{2}\right)\left(\ket{00}+\ket{11}\right)$,
where $\ket{m_{A}m_{B}}=\ket{m_{A}}\otimes\ket{m_{B}}$ denotes a
product of $A$ and $B$ computational basis states and $m_{A},m_{B}\in\left\{ 0,1\right\} $.
It is assumed that $A$ and $B$ are separated spatially so that Coulomb
interactions between them are negligible, but that each MCQ interacts
with its own local environment. This separation could be established
after preparation in $\ket{\Psi_{AB}\left(0\right)}$, or some remote
entanglement mechanism could be applied after separation. The dynamics
of the loss of entanglement in $AB$\textemdash not the means of entanglement\textemdash are
the focus of this work.

\subsection{The Environment}

\begin{figure}
\begin{centering}
\includegraphics[scale=0.6]{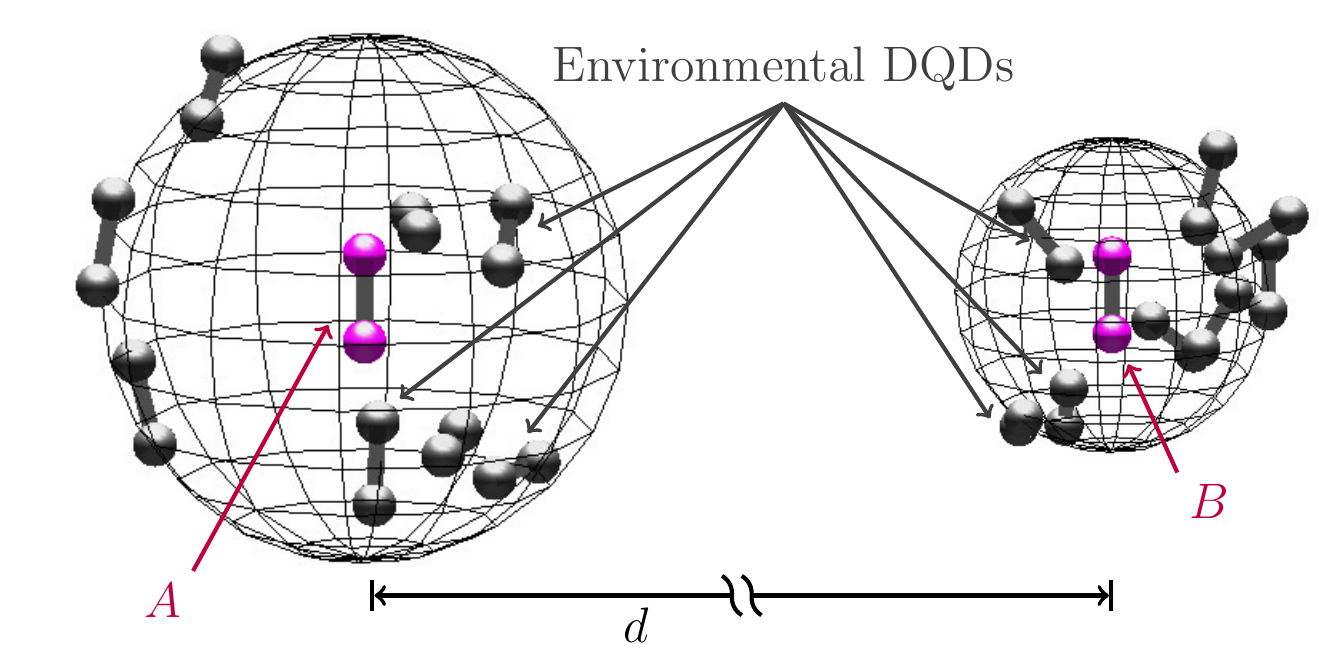}
\par\end{centering}
\caption{The target qubits, $A$ and $B$, are entangled and coupled to local
environments comprised of randomly-oriented DQDs. Colored spheres
represent molecular quantum dots, and a connecting bar indicates the
intramolecular tunneling path. The MCQs of the target pair are marked
with purple-colored dots. There are $M$ environmental molecules distributed
randomly and with random orientations about the surface of a sphere
of radius $R_{X}$ for qubit $X\in\{A,B\}$. The target pair $AB$
is entangled over a large distance $d\gg R_{A},R_{B}$ so that MCQ
$A$ and its environment have negligible electrostatic interactions
with MCQ $B$ and its environment.\label{fig:cell_visual_differ}}
\end{figure}

The local environment for each MCQ in $AB$ is explicitly modeled
using $M$ DQDs surrounding each target MCQ.\footnote{The MCQs and the environmental molecules all are assumed to be DQDs
of the same molecular species. However, for clarity, ``MCQ'' is
reserved for the target pair of DQDs used to model qubits; on the
other hand, ``DQD'' is more general and may be applied to both target
molecules and environmental molecules. Following this train of thought,
we reserve the term ``computational basis'' to describe fully-localized
electronic states of the MCQs in $AB$, but the term ``classical
basis'' could describe an analogous state in any system of DQDs\textendash either
MCQ or environmental.} The $M$ environmental DQDs are arranged on the surface of a sphere
of radius $R_{X}$ centered on qubit $X\in\{A,B\}$, as depicted in
Figure \ref{fig:cell_visual_differ}. Here, the orientations and positions
on the sphere of the environmental molecules are randomized. Generally,
$R_{A}\neq R_{B}$ so that one MCQ in $AB$ may have a stronger environmental
interaction than does its partner. This generalizes a previous study,
in which $R_{A}=R_{B}$ was a constraint,\citep{blair2018entanglement}
so that neither MCQ suffered the dominant environmental interaction.
We designate the two local environments together as the complete environment,
$\mathcal{E}$, with $N=2M$ environmental DQDs.
Environmental product states may be formed by taking tensor products
\begin{equation}
\ket{\vec{m}_{p}}=\ket{m_{N}}\ket{m_{N-1}}\cdots\ket{m_{k}}\cdots\ket{m_{2}}\ket{m_{1}}\;,\label{eq:environmental_state}
\end{equation}
where a counting number, $k$, indexes the environmental DQDs, and
$m_{k}\in\{0,1\}$ labels a classical basis state for the $k$-th
environmental molecule. The $N$-element binary vector,
\begin{equation}
\vec{m}_{p}=m_{N}m_{N-1}\cdots m_{k}\cdots m_{2}m_{1},\label{eq:env_binary_word}
\end{equation}
then, specifies an environmental product state, and $p\in\{0,1,2,\ldots,2^{N}-1\}$
is a whole-number representation of $\vec{m}_{p}$.

In this paper, the initial state of the environment, $\ket{\mc{E}\Argum{0}}$,
is a product state of environmental DQDs, each prepared in a superposition
$\ket{\psi_{k}}=(1/\sqrt{2})\left(\ket{0}+e^{i\phi_{k}}\ket{1}\right)$:
\[
\ket{\mathcal{E}\Argum{0}}=\ket{\psi_{N}}\ket{\psi_{N-1}}\cdots\ket{\psi_{2}}\ket{\psi_{1}}.\;
\]
Here, the relative phase, $\phi_{k}$, is randomly selected. In the
environmental classical basis, $\{\ket{\vec{m}_{p}}\}$, the state
$\ket{\mathcal{E}\Argum{0}}$ may be written as
\begin{equation}
\ket{\mc{E}\Argum{0}}=\frac{1}{2^{N/2}}\sum_{\vec{m}_{p}}e^{i\Phi\Argum{\vec{m}_{p}}}\;\ket{\vec{m}_{p}}.\;\label{eq:InitialEnvironmentalState}
\end{equation}
where
\[
\Phi\Argum{\vec{m}_{p}}=\sum_{k=1}^{N}\left[\vec{m}_{p}\right]_{k}\phi_{k}\;,
\]
and $\left[\vec{m}_{p}\right]_{k}$ denotes the $k$-th bit of $\vec{m}_{p}$.

\subsection{System Dynamics}

The Hamiltonian of the global system, $\Omega\equiv AB\mc{E}$, is
determined by the Coulomb interactions between all the DQDs of $\Omega$.
Let $U_{m_{j},m_{k}}^{j,k}$ be the electronstatic potential energy
between the $j$-th DQD in state $m_{j}$ and the $k$-th DQD in state
$m_{k}$. This energy is given by
\begin{equation}
U_{m_{j},m_{k}}^{j,k}=\frac{P\left(m_{j}\right)P\left(m_{k}\right)e^{2}}{16\pi\epsilon_{0}}\left[\frac{1}{r_{0,0}^{j,k}}-\frac{1}{r_{0,1}^{j,k}}-\frac{1}{r_{1,0}^{j,k}}+\frac{1}{r_{1,1}^{j,k}}\right]\;,\label{eq:coulomb}
\end{equation}
where $\epsilon_{0}$ is the permittivity of free space; $r_{m_{j},m_{k}}^{j,k}$is
the distance between dot $m_{j}$ in DQD $j$ and dot $m_{k}$ in
DQD $k$; $P(m)$ is polarization of a DQD in state $m$; and $P(1)=+1$
and $P(0)=-1$.

Let $E_{m_{A}m_{B}}\left(\vec{m}_{p}\right)$ be the total electrostatic
potential energy of a global state $\Ket{\Phi_{m_{A},m_{B};\vec{m}_{p}}}$
defined as
\[
\Ket{\Phi_{m_{A},m_{B};\vec{m}_{p}}}=\ket{m_{A}m_{B}}\otimes\Ket{\vec{m}_{p}}\;.
\]
The energy $E_{m_{A}m_{B}}\left(\vec{m}_{p}\right)$ is calculated
by summing over all DQD pair-wise interactions in $\Omega$:
\begin{align}
E_{m_{A}m_{B}}\left(\vec{m}_{p}\right) & =\Braket{\Phi_{m_{A},m_{B};\vec{m}_{p}}|\hat{H}|\Phi_{m_{A},m_{B};\vec{m}_{p}}}\nonumber \\
 & =\frac{1}{2}\underset{j\neq k}{\sum}U_{m_{j},m_{k}}^{j,k}\label{eq:total_electrostatic-1}
\end{align}
Here, $\hat{H}$ is the Hamiltonian for $\Omega$, and the indices
of summation, $i$ and $j$, include each DQD in $\Omega$: $i,j\in\left\{ A,B,1,2,\ldots,N\right\} $.

To eliminate complicating dissipative effects, this study of disentanglement
is constrained to the regime where tunneling between states $\ket{0}$
and $\ket{1}$ is suppressed. In this limit, the global Hamiltonian
may be written as
\begin{align*}
\hat{H} & =\sum_{m_{A,}m_{B}}\ket{m_{A}m_{B}}\bra{m_{A}m_{B}}\\
 & \qquad\otimes\sum_{\vec{m}_{p}}E_{m_{A}m_{B}}\left(\vec{m}_{p}\right)\ket{\vec{m}_{p}}\bra{\vec{m}_{p}}\;.
\end{align*}
The Hamiltonian is diagonal in the global basis $\{\ket{\Phi_{m_{A}m_{B};\vec{m}_{p}}}\}$.

\subsubsection{\label{subsect:GlobalDynamics}Global System Dynamics}

The dynamics of the global system are described exactly within this
model using the Schrödinger equation,
\[
\frac{\partial}{\partial t}\ket{\Psi\left(t\right)}=-\frac{i}{\hbar}\hat{H}\ket{\Psi\left(t\right)}\,.
\]
The time-dependent state, $\ket{\Psi\left(t\right)}$, is obtained
by applying the time evolution operator, $\hat{U}\left(t\right)=\exp\,(-i\hat{H}t/\hbar)$
to the initial state $\ket{\Psi\left(0\right)}$:
\begin{equation}
\ket{\Psi\left(t\right)}=\hat{U}\left(t\right)\ket{\Psi\left(t\right)}\;.\label{eq:TimeDependentState_TDSE_Soln}
\end{equation}

\subsubsection{\label{subsec:ReducedDynamics}Reduced Dynamics of the Target MCQ
Pair}

Unlike the initial state, the time-dependent $\ket{\Psi\left(t\right)}$
generally is not a product of an $AB$ state $\ket{\Psi_{AB}\left(t\right)}$
and an environmental state $\ket{\mc{E}\left(t\right)}$ . This is
due to the interaction between $AB$ and $\mc{E}$, which causes entanglement
between $AB$ and $\mc{E}$ over time, as well as the unraveling of
$A$-$B$ entanglement.

While $AB$ may no longer have its own local state for $t>0$, the
best time-dependent, local description possible for $AB$ is its reduced
density matrix, $\hat{\rho}_{AB}^{(r)}(t)$. This is obtained by forming
the time-dependent global density matrix, $\hat{\rho}_{\Omega}(t)=\Ket{\Psi\left(t\right)}\Bra{\Psi\left(t\right)}$,
and tracing $\hat{\rho}_{\Omega}(t)$ over the environmental degrees
of freedom:
\begin{equation}
\hat{\rho}_{AB}^{(r)}(t)=\Tr_{\mathcal{E}}\left(\hat{\rho}_{\varOmega}(t)\right)=\sum_{j_{\mathcal{E}}}\Braket{j_{\mathcal{E}}|\hat{\rho}_{\varOmega}|j_{\mathcal{E}}}\;.\label{eq:reduced-density}
\end{equation}
Here, $\Tr_{\mathcal{E}}$ denotes the trace over the degrees of freedom
of $\mathcal{E}$, and $\left\{ \ket{j_{\mathcal{E}}}\right\} $ is
any orthonormal basis for the $\mc{E}$. Henceforth, we drop the superscript
$\left(r\right)$ from the reduced density matrix for $AB$.

This model is designated as the ``numerical'' model, in which the
dynamics of $AB$ and $\mc{E}$ are calculated explicitly in order
to obtain $\hat{\rho}_{AB}\left(t\right)$.

\subsection{Semi-analytic Model}

Here, an analytical treatment is used to find $\hat{\rho}_{AB}(t)$
without explicitly calculating the dynamics of $\mc{E}$.

The initial state vector for the system and environment is a product
of the system and environment initial states from Equations (\ref{eq:BellState})
and (\ref{eq:InitialEnvironmentalState}):
\begin{align}
\ket{\Psi\left(0\right)} & =\frac{1}{\sqrt{2}}\left(\ket{00}+\ket{11}\right)\otimes\frac{1}{2^{N/2}}\sum_{\vec{m}_{p}}e^{i\Phi\left(\vec{m}_{p}\right)}\ket{\vec{m}_{p}}.\label{eq:InitialStateExplicit}
\end{align}
Because $\hat{H}$ is diagonal in the global basis $\{\ket{\Phi_{m_{A}m_{B};\vec{m}_{p}}}\}$,
so also is the time evolution operator, $\hat{U}\Argum{t}$:
\begin{align}
\hat{U}\Argum{t} & =\sum_{m_{A},m_{B}}\ket{m_{A}m_{B}}\bra{m_{A}m_{B}}\nonumber \\
 & \qquad\otimes\sum_{\vec{m}_{p}}e^{-iE_{m_{A}m_{B}}\Argum{\vec{m}_{p}}t/\hbar}\ket{\vec{m}_{p}}\bra{\vec{m}_{p}}.\label{eq:TimeEvOp}
\end{align}
Thus, the time-dependent global state $\ket{\Psi\Argum{t}}$ is found
by using Equations (\ref{eq:TimeDependentState_TDSE_Soln}), (\ref{eq:InitialStateExplicit})
and (\ref{eq:TimeEvOp}):
\begin{align*}
\ket{\Psi\Argum{t}} & =\frac{1}{2^{\left(N+1\right)/2}}\ket{00}\otimes\sum_{\vec{m}_{p}}e^{-iE_{00}\left(\vec{m}_{p}\right)t/\hbar}e^{i\Phi\left(\vec{m}_{p}\right)}\ket{\vec{m}_{p}}\\
 & \quad+\frac{1}{2^{\left(N+1\right)/2}}\ket{11}\otimes\sum_{\vec{m}_{p}}e^{-iE_{11}\left(\vec{m}_{p}\right)t/\hbar}e^{i\Phi\left(\vec{m}_{p}\right)}\ket{\vec{m}_{p}}\;.
\end{align*}
This may be used to form the global $\hat{\rho}_{\Omega}(t)$, which,
when traced over the classical environmental basis, $\left\{ \ket{\vec{m}_{p}}\right\} $,
yields the reduced density matrix for the target MCQ pair:
\begin{align}
\hat{\rho}_{AB}\Argum{t} & =\frac{1}{2}\left(\ket{00}\bra{00}+\ket{11}\bra{11}\right)\nonumber \\
 & \quad+\frac{1}{2^{N+1}}\ket{00}\bra{11}\sum_{\vec{m}_{p}}e^{-i\omega\TextSub{flip}\Argum{\vec{m}_{p}}t}\nonumber \\
 & \quad+\frac{1}{2^{N+1}}\ket{11}\bra{00}\sum_{\vec{m}_{p}}e^{i\omega\TextSub{flip}\Argum{\vec{m}_{p}}t}\;.\label{eq:TimeDependentRDMAB}
\end{align}
Here, we have defined the double-bit-flip frequency
\begin{equation}
\omega\TextSub{flip}\Argum{\vec{m}_{p}}\equiv\frac{1}{\hbar}\left(E_{11}\Argum{\vec{m}_{p}}-E_{00}\Argum{\vec{m}_{p}}\right)=\frac{1}{\hbar}E\TextSup{flip}_{\vec{m}_{p}}\;,\label{eq:omega_defined}
\end{equation}
which is proportional to the double-bit-flip energy
\begin{equation}
E\TextSup{flip}_{\vec{m}_{p}}=E_{11}\Argum{\vec{m}_{p}}-E_{00}\Argum{\vec{m}_{p}},\label{eq:DoubleBitFlipEnergy}
\end{equation}
the cost of a double bit flip of $AB$ given enviromental state $\ket{\vec{m}_{p}}$.

We designate the model of Equation (\ref{eq:TimeDependentRDMAB})
a ``semi-analytic'' model, since an analytic treatment was used
to obtain Equation (\ref{eq:TimeDependentRDMAB}), but the numerous
energies, $\{E\TextSup{flip}_{\vec{m}_{p}}\}$, and $\hat{\rho}_{AB}\left(t\right)$
must be evaluated numerically. This model alleviates the significant
burden of explicitly calculating the dynamics of $\mc{E}$.

\subsection{Approximate Gaussian Model}

Now, consider the summations in Equation (\ref{eq:TimeDependentRDMAB}).
Together with the factor $1/2^{N}$, these may be written as
\begin{align}
\frac{1}{2^{N}}\sum_{\vec{m}_{p}}e^{\pm i\omega\TextSub{flip}\Argum{\vec{m}_{p}}t} & =\sum_{k}\frac{1}{k!}\left(\pm it\right)^{k}\frac{1}{2^{N}}\sum_{\vec{m}_{p}}\omega\TextSub{flip}^{k}\left(\vec{m}_{p}\right)\nonumber \\
 & =\sum_{k}\frac{1}{k!}\left(\pm it\right)^{k}\Braket{\omega^{k}},\label{eq:SumOfAverages}
\end{align}
where we define
\[
\Braket{\omega^{k}}\equiv\frac{1}{2^{N}}\sum_{\vec{m}_{p}}\omega\TextSub{flip}^{k}\left(\vec{m}_{p}\right)\;.
\]
Here, $\braket{\omega^{k}}$ is an average over $\{\omega\TextSub{flip}^{k}\left(\vec{m}_{p}\right)\}$,
and we identify $\Braket{\omega^{1}}=\bar{\omega}$ and $\sqrt{\Braket{\omega^{2}}}=\omega\TextSub{RMS}\TextSup{flip}$
as average and root-mean-square values, respectively, of the frequencies
$\{\omega\TextSub{flip}\Argum{\vec{m}_{p}}\}$. Similarly, we can
define averages of the double-bit-flip energies and their powers:
\[
\Braket{\left(E\TextSup{flip}_{\vec{m}_{p}}\right)^{k}}\equiv\frac{1}{2^{N}}\sum_{\vec{m}_{p}}\left(E\TextSup{flip}_{\vec{m}_{p}}\right)^{k}=\hbar^{n}\Braket{\omega^{k}}\;,
\]
with a mean double-bit-flip energy,
\begin{equation}
\bar{E}\TextSup{flip}_{\vec{m}_{p}}=\Braket{(E\TextSup{flip}_{\vec{m}_{p}})^{1}}=\hbar\bar{\omega}\;,\label{eq:MeanBitFlipEnergy}
\end{equation}
 and a root-mean-square double-bit-flip energy,
\begin{equation}
\sigma_{E}=E\TextSup{flip}\TextSub{RMS}=\sqrt{\Braket{(E\TextSup{flip}_{\vec{m}_{p}})^{2}}}=\hbar\omega\TextSub{RMS}\TextSup{flip}\;.\label{eq:EFlipRMS}
\end{equation}

For environments with randomly-placed and randomly-oriented DQDs\textemdash more
generally than just the spherical environments modeled in this paper\textemdash the
frequencies $\{\omega\TextSub{flip}\left(\vec{m}_{p}\right)\}$ and
energies $\{E\TextSup{flip}_{\vec{m}_{p}}(\vec{m}_{p})\}$ will tend
to be normally distributed. Thus, on average, a random environment
will have small $\Braket{\omega^{k}}$ and small $\braket{(E\TextSup{flip}_{\vec{m}_{p}}(\vec{m}_{p}))^{k}}$
for odd $k$. Neglecting these terms from Equation (\ref{eq:SumOfAverages})
as well as terms beyond the third order in $t$, we have the approximation
\begin{align}
\frac{1}{2^{N}}\sum_{\vec{m}_{p}}e^{\pm i\omega\TextSub{flip}\Argum{\vec{m}_{p}}t} & =1\pm it\braket{\omega^{1}}-\frac{t^{2}}{2}\Braket{\omega^{2}}\nonumber \\
 & \quad\mp i\frac{t^{3}}{3!}\Braket{\omega^{3}}+\cdots\nonumber \\
 & \simeq1-\frac{\left(\omega\TextSub{RMS}\TextSup{flip}t\right)^{2}}{2}\nonumber \\
 & \simeq e^{-\left(\omega\TextSub{RMS}\TextSup{flip}\right)^{2}t^{2}/2}\;.\label{eq:GaussianApproximation}
\end{align}
Now, inserting Equation (\ref{eq:GaussianApproximation}) into $\hat{\rho}_{AB}\left(t\right)$
of Equation (\ref{eq:TimeDependentRDMAB}), the coherences $\Braket{00|\hat{\rho}_{AB}|11}=\Braket{11|\hat{\rho}_{AB}|00}^{\ast}$
have a time-dependence with a Gaussian decay:
\begin{align}
\hat{\rho}_{AB}\Argum{t} & \simeq\frac{1}{2}\bigg[\ket{00}\bra{00}+\ket{11}\bra{11}\nonumber \\
 & \qquad+e^{-\left(\omega\TextSub{RMS}\TextSup{flip}\right)^{2}t^{2}/2}\left(\ket{00}\bra{11}+\ket{11}\bra{00}\right)\bigg]\label{eq:approximate_RDMAB}
\end{align}

The main assumption behind the Gaussian approximate model for $\hat{\rho}_{AB}\Argum{t}$
is randomness in the environment. The Gaussian model could be applied
more broadly to describe the dynamics of disentanglement due to other
environments where randomness is a feature, as well.

\subsubsection{Application to Local Spherical Environments}

\begin{figure*}
\begin{centering}
\subfloat[]{\begin{centering}
\includegraphics[width=0.475\textwidth,trim={0.125cm 0 1.125cm 0}, clip]{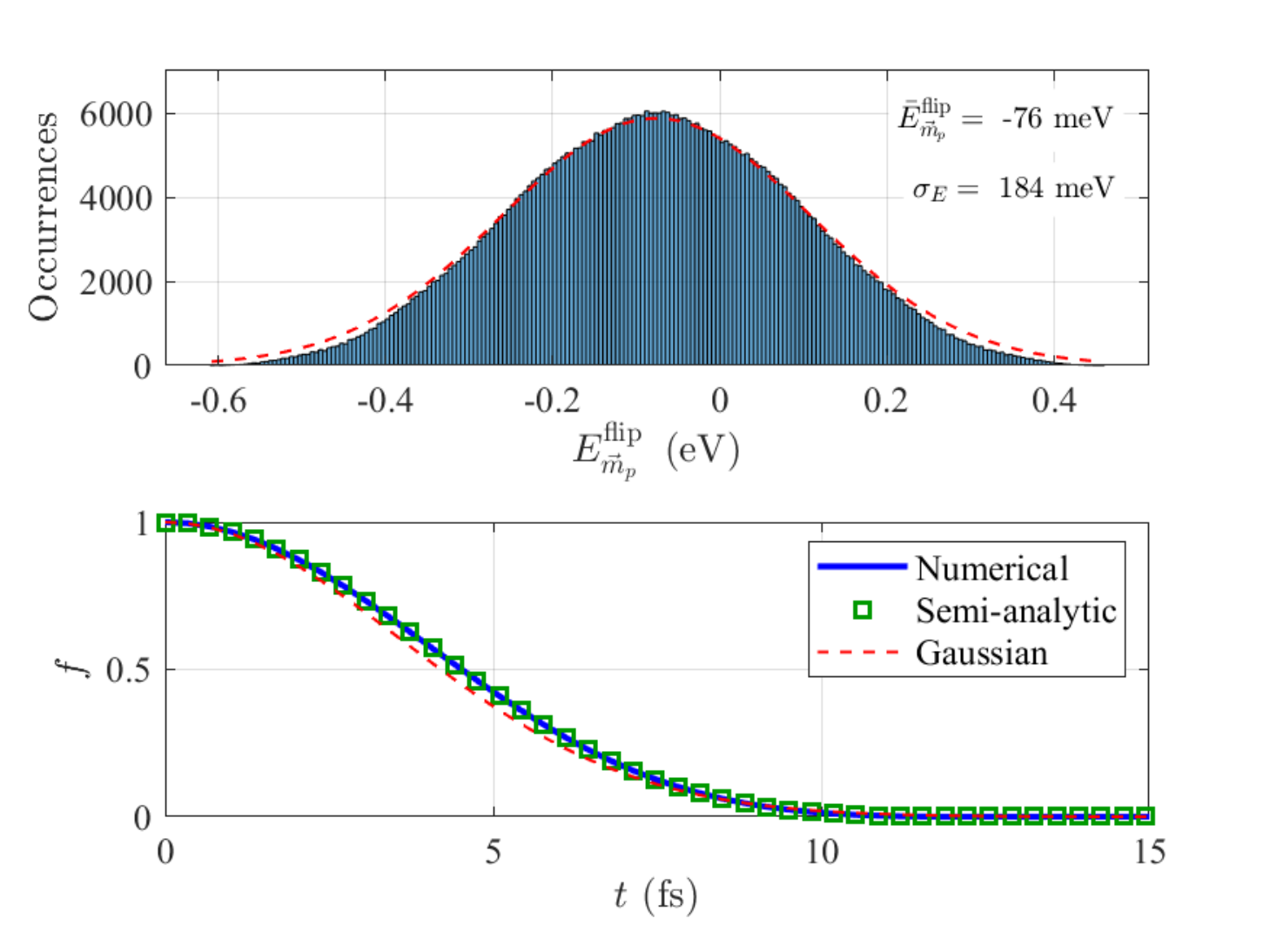}
\par\end{centering}
}\hfill{}\subfloat[]{\begin{centering}
\includegraphics[width=0.475\textwidth,trim={0.125cm 0 1.125cm 0}, clip]{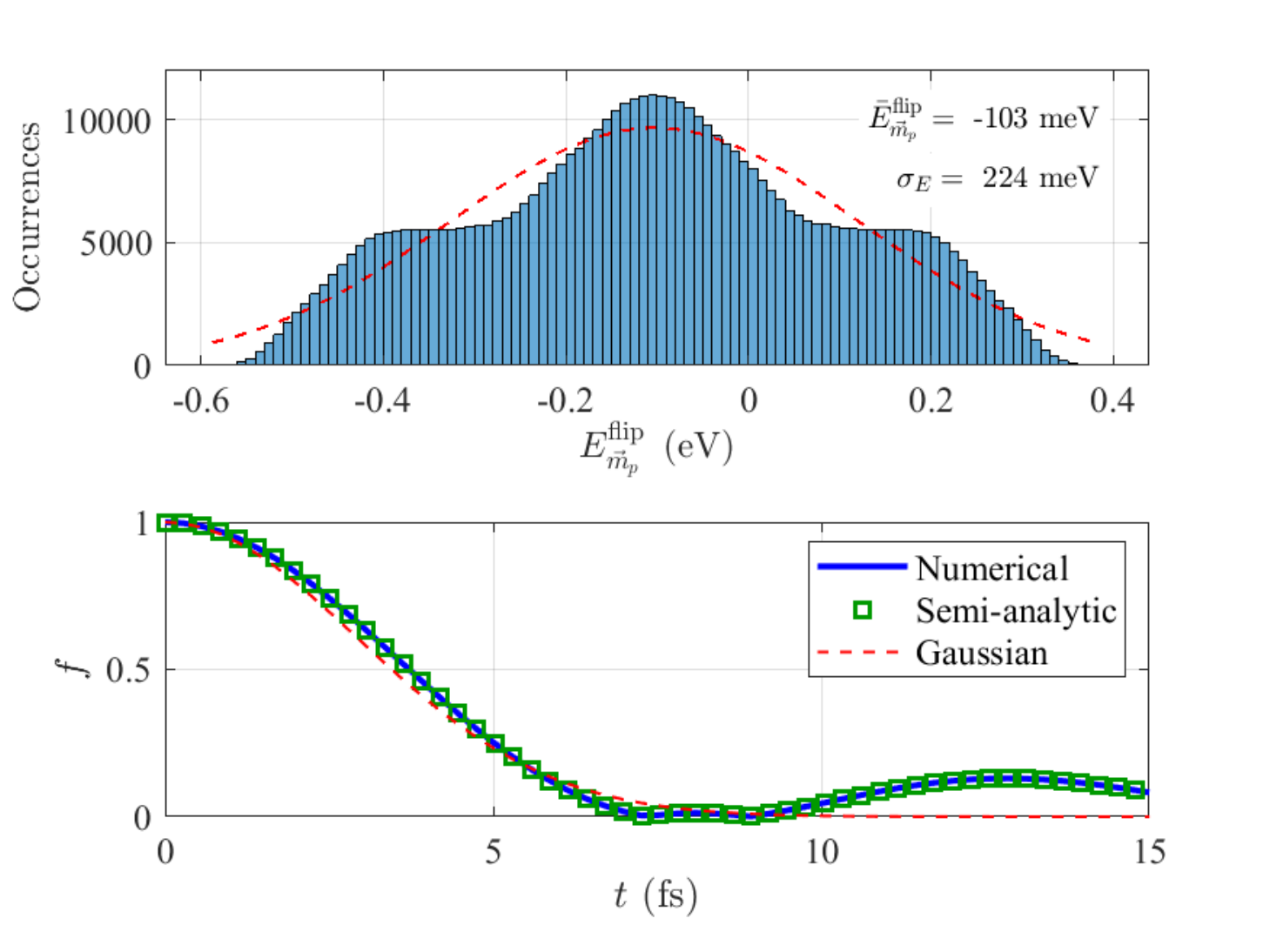}
\par\end{centering}
}
\par\end{centering}
\caption{As the distribution of double-bit-flip energies $\{E\TextSup{flip}_{\vec{m}_{p}}\}$
approaches a zero-centric Gaussian distribution, the time dependence
of the decay in coherences $\Braket{00|\hat{\rho}_{AB}|11}=\Braket{11|\hat{\rho}_{AB}|00}^{\ast}$
becomes more Gaussian. (a) A historgram of the energies $\{E\TextSup{flip}_{\vec{m}_{p}}\}$
(upper plot) which approaches an ideal Gaussian (dashed red line)
centered at the origin corresponds to a highly-Gaussian form in the
decay of coherences. (b) Deviations from a zero-centric Gaussian distribution
in energies $\{E\TextSup{flip}_{\vec{m}_{p}}\}$ introduces non-Gaussian
behavior in the decay of $f$. In both cases shown, $a=1$ nm, the
environmental radii are $R_{A}/R_{B}=4a/2a$, and the environmental
populations are $N=20$. \label{fig:Eflip_and_TimeDependence}}
\end{figure*}

The double-sphere environments studied in this context provide a concrete
example of this analysis. Here, a histogram of the energies $\{E\TextSup{flip}_{\vec{m}_{p}}\}$
is plotted for a particular random environment. To provide a qualitative
visual cue for how Gaussian the distribution is, a fitting function,
\[
g\left(E\TextSup{flip}_{\vec{m}_{p}}\right)=Ae^{-\left(E\TextSup{flip}_{\vec{m}_{p}}-\bar{E}\TextSup{flip}_{\vec{m}_{p}}\right)^{2}/2\sigma_{E}^{2}},
\]
also is plotted (dashed red curve), where $A$ is chosen to minimize
curve-fitting error. The highly-Gaussian energy distribution shown
in the upper panel results in a highly-Gaussian time-dependence for
the magnitude of the coherences of $\hat{\rho}_{AB}$, shown in the
lower panel of \ref{fig:Eflip_and_TimeDependence}(a). Here, the ratio
$f\left(t\right)$ is plotted, which is defined as the magnitude of
non-zero coherences relative to their initial magnitudes:
\begin{equation}
f\left(t\right)\equiv\frac{\left|\Braket{00|\hat{\rho}_{AB}\left(t\right)|11}\right|}{\left|\Braket{00|\hat{\rho}_{AB}\left(0\right)|11}\right|}=\frac{\left|\Braket{11|\hat{\rho}_{AB}\left(t\right)|00}\right|}{\left|\Braket{11|\hat{\rho}_{AB}\left(0\right)|00}\right|}\;.\label{eq:CoherenceRatio}
\end{equation}
In the plot of $f\left(t\right)$, the approximate Gaussian decay
from Equation (\ref{eq:approximate_RDMAB}) is shown using a dashed
red line, and deviations from this approximate behavior are attributed
to the terms neglected from Equation (\ref{eq:GaussianApproximation}).

Figure \ref{fig:Eflip_and_TimeDependence}(b) provides an example
of an environment in which the distribution $\{E\TextSup{flip}_{\vec{m}_{p}}\}$
deviates from a Gaussian form (upper panel). Here, a larger $\bar{E}\TextSup{flip}_{\vec{m}_{p}}$
leads to a larger $\bar{\omega}$; and , other terms for higher odd
powers of $t$ neglected in Equation (\ref{eq:approximate_RDMAB})
introduce non-zero imaginary components which drive departures from
a purely Gaussian time dependence in the coherences of $\hat{\rho}_{AB}\left(t\right)$.
Thus, more notable deviations from the red Gaussian line appear in
the corresponding plot of $f\left(t\right)$ of the lower panel.

Henceforth, we refrain from calculating results using the fully numerical
model, since it is more computationally-intensive than the semi-analytic
treatment. This is justified, since the lower panels of subfigures
\ref{fig:Eflip_and_TimeDependence}(a) and \ref{fig:Eflip_and_TimeDependence}(b)
demonstrate exact agreement between the numerical and semi-analytic
models.

\subsection{A Time Scale for Disentanglement}

Let the decay of non-zero coherences in Equation (\ref{eq:approximate_RDMAB})
be mapped to a Gaussian with standard deviation $\sigma_{t}$, $g\left(t\right)\propto\exp\left(-\left(t-t_{0}\right)^{2}/2\sigma_{t}^{2}\right)$.
Then, for this decay, $t_{0}=0$ and $\sigma_{t}=1/\omega\TextSub{RMS}\TextSup{flip}$.
Thus, the root-mean-square double-bit-flip frequency characterizes
the Gaussian decay of the coherences of $\hat{\rho}_{AB}\left(t\right)$.
We define 
\begin{equation}
\tau_{E}=\frac{\pi}{\omega\TextSub{RMS}\TextSup{flip}}=\frac{\pi\hbar}{E\TextSup{flip}\TextSub{RMS}} \; .\label{eq:TimeScaleTauE}
\end{equation}
as a time scale for the dynamics of disentanglment. Here, the factor
of $\pi$ is included to make $\tau_{E}$ directly comparable to $\tau$,
the time scale from previous work.\citep{blair2018entanglement}

\subsection{\label{subsubsect:measures_of_ent}Measures of Entanglement}

To quantify entanglement between $A$ and $B$, we use three correlation
functions: $S\TextSub{BM}$, the Bell-Mermin (BM) correlation function;\citep{Mermin1985}
$S\TextSub{CHSH}$, the Clauser-Horne-Shimony-Holt (CHSH) correlation
function;\citep{clauser1969proposed} and $S\TextSub{BPRV}$, the
Brukner-Paunkovi\'c-Rudolph-Vedral (BPRV) correlation function.\citep{brukner2006entanglement}
These are functions of the two-qubit reduced density matrix $\hat{\rho}_{AB}(t)$.
The details of our implementations of the correlation functions are
discussed either here or in the previous work by Blair, Tóth, and
Lent.\citep{blair2018entanglement}

\subsubsection{The Bell-Mermin Correlation Function}

The Bell-Mermin correlation used here is formulated for two qubits,
$A$ and $B$, measured independently with three measurement settings,\citep{Mermin1985}
$j\in\left\{ 1,2,3\right\} $ corresponding to three rotated angles
of measurement $\left\{ \theta_{j}\right\} $. The Bell-Mermin correlation
function, $S\TextSub{BM}$, is
\begin{equation}
S\TextSub{BM}=\text{Tr}\left(\hat{\rho}\hat{P}\TextSub{same}\right),\label{eq:BellMermin}
\end{equation}
where
\begin{align*}
\hat{P}\TextSub{same} & \equiv\sum_{i=1,j\neq i}^{3}\sum_{m=0}^{1}\hat{R}\left(\theta_{i}\right)\ket{m}\bra{m}\hat{R}\left(-\theta_{i}\right)\\
 & \qquad\qquad\otimes\hat{R}\left(\theta_{j}\right)\ket{m}\bra{m}\hat{R}\left(-\theta_{j}\right),
\end{align*}
and $R\left(\theta\right)$ is a single-qubit rotation operator:
\[
R\left(\theta\right)=\cos\theta\left(\ket{0}\bra{0}+\ket{1}\bra{1}\right)+\sin\theta\left(\ket{0}\bra{1}-\ket{1}\bra{0}\right)\;.
\]
$S\TextSub{BM}$ may be interpreted as the sum of the probabilities
that a measurement on each MCQ will yield the same result, 0 or 1,
when measured in dissimilar bases. A value of $S\TextSub{BM}\leq1$
is not possible for a pair of particles described by purely classical
statistics assuming local realism, so this is designated the ``Bell
violation'' regime. To maximize the Bell violation of measurements
on $\hat{\rho}_{AB}\left(t\right)$, we choose $\left(\theta_{1},\theta_{2},\theta_{3}\right)=\left(0,\pi/3,2\pi/3\right)$.

Applying the exact, semi-analytic $\hat{\rho}_{AB}$ of Equation (\ref{eq:TimeDependentRDMAB})
to Equation (\ref{eq:BellMermin}), we obtain

\[
S\TextSub{BM}\left(t\right)=\frac{9}{8}-\frac{3}{8}\frac{1}{2^{N}}\sum_{\vec{m}_{p}}\cos\left[\omega\left(\vec{m}_{p}\right)t\right].
\]
The approximate $\hat{\rho}_{AB}$ of Equation (\ref{eq:approximate_RDMAB})
leads to
\[
S\TextSub{BM}\left(t\right)\simeq\frac{9}{8}-\frac{3}{8}e^{-\omega^{2}\TextSub{RMS}t^{2}/2}.
\]
The approximate form of $S\TextSub{BM}$ clearly highlights the initial
and asymptotic values of $S\TextSub{BM}\left(t\right)$: $S\TextSub{BM}\left(0\right)=3/4$,
and $S\TextSub{BM}\left(\infty\right)=9/8.$ Thus, the pair $AB$
starts maximally-entangled in the Bell violation regime, and time
evolution unravels this entanglement through interaction and entanglement
with $\mc{E}$.

\subsubsection{The Clauser-Horne-Shimony-Holt (CHSH) Correlation Function}

Similarly, the CHSH correlation function as implemented by Blair,
Tóth, and Lent\citep{blair2018entanglement} may be applied to the
semi-analytic version of $\hat{\rho}_{AB}\left(t\right)$ of Equation
(\ref{eq:TimeDependentRDMAB}), with result
\[
S\TextSub{CHSH}\left(t\right)=\sqrt{2}\left|1+\frac{1}{2^{N}}\sum_{\vec{m}_{p}}\cos\left[\omega\left(\vec{m}_{p}\right)t\right]\right|;
\]
or the approximate $\hat{\rho}_{AB}\left(t\right)$ of Equation (\ref{eq:approximate_RDMAB}),
leading to
\[
S\TextSub{CHSH}\left(t\right)\simeq\sqrt{2}\left|1+e^{-\omega^{2}\TextSub{RMS}t^{2}/2}\right|.
\]
Here, the Bell violation regime is $S\TextSub{CHSH}>2$. By this measure
of entanglement, the $AB$ pair starts well within the Bell violation
region with $S\TextSub{CHSH}\left(0\right)=2\sqrt{2}$, but $AB$
eventually crosses out to a classically-describable region with $S\TextSub{CHSH}\left(\infty\right)=\sqrt{2}$.

\subsubsection{The Brukner-Paunkovi\'c-Rudolph-Vedral (BPRV) Correlation Function}

Finally, the BPRV correlation is calculated for the exact $\hat{\rho}_{AB}\left(t\right)$
of Equation (\ref{eq:TimeDependentRDMAB}) as
\begin{equation}
S\TextSub{BPRV}\left(t\right)=6+\frac{3}{2^{N+1}}\sum_{\vec{m}_{p}}\cos\left[\omega\left(\vec{m}_{p}\right)t\right].\label{eq:SBPRV_exact}
\end{equation}
The approximate $\hat{\rho}_{AB}\left(t\right)$ of Equation (\ref{eq:approximate_RDMAB})
results in
\begin{equation}
S\TextSub{BPRV}\left(t\right)\simeq6+\frac{3}{2}e^{-\omega\TextSub{RMS}^{2}t^{2}/2}.\label{eq:SBPRV_approximate}
\end{equation}
The details of our $S\TextSub{BPRV}$ calculation are found in previous
work.\citep{blair2018entanglement} Here, the Bell violation regime
is defined by $S\TextSub{BPRV}>7$. Initially maximally entangled,
$AB$ has $S\TextSub{BPRV}\left(0\right)=15/2$, and time evolution
brings $AB$ out of the Bell violation regime to an asymptotic value
of $S\TextSub{BPRV}\left(\infty\right)=6$.


\section{Results\label{sec:Results}}

\subsection{Validation of $\tau_{E}$ as a Time Scale}

Part of the motivation for this work was that $\tau=\sqrt{\tau_{A}\tau_{B}}$,
a previous disentanglement time scale\citep{blair2018entanglement}
used in the case where $R_{A}=R_{B}$, did not generalize well to
cases in which $R_{A}\neq R_{B}$. Here, $\tau_{A}$ and $\tau_{B}$
are time scales for the decoherence of each single qubit within its
own local environment.\citep{RamseyKrausOp2017}

\begin{figure*}
\begin{centering}
\subfloat[]{\centering{}\includegraphics[width=0.5\textwidth]{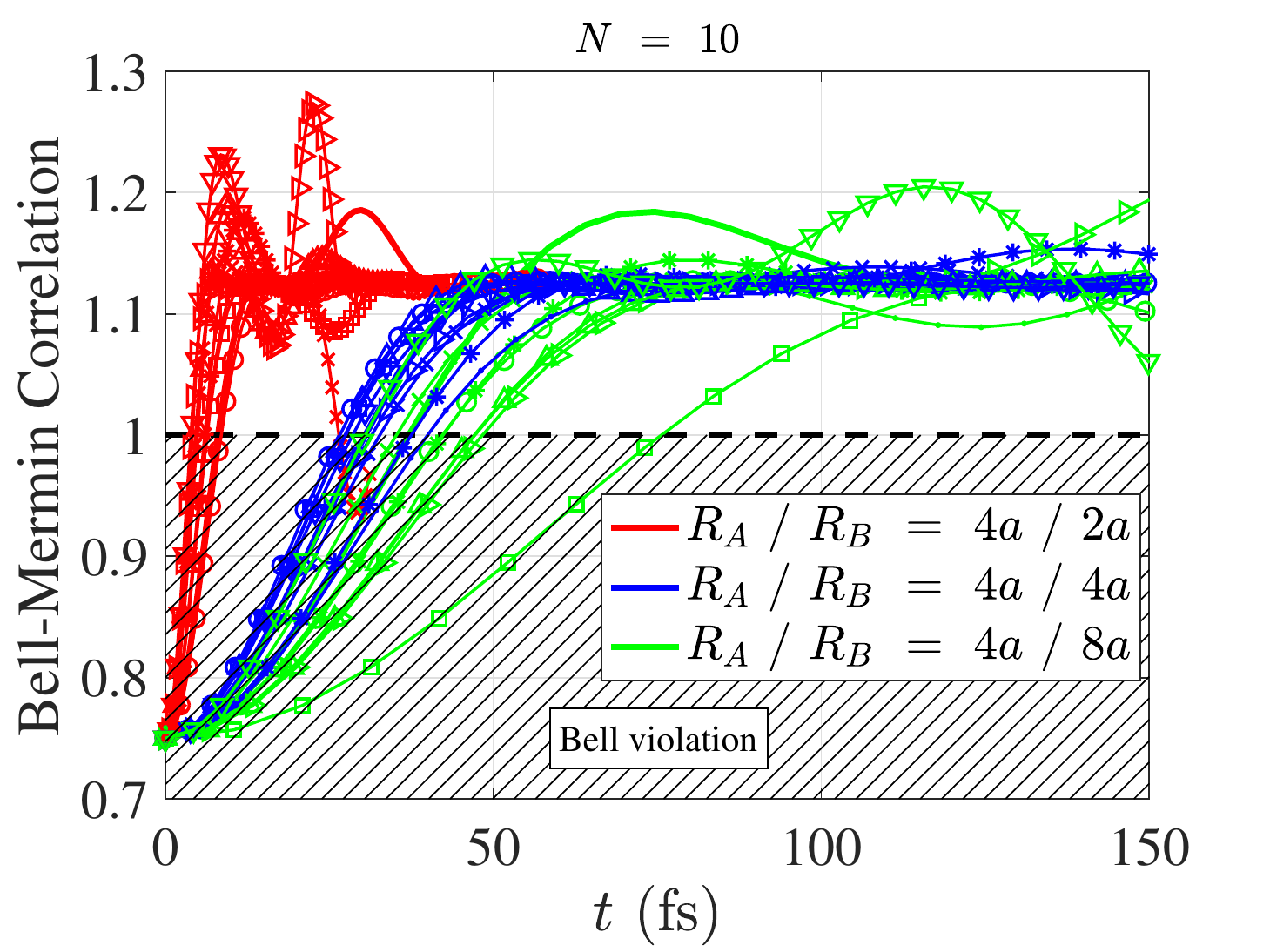}}\hfill{}\subfloat[]{\centering{}\includegraphics[width=0.5\textwidth]{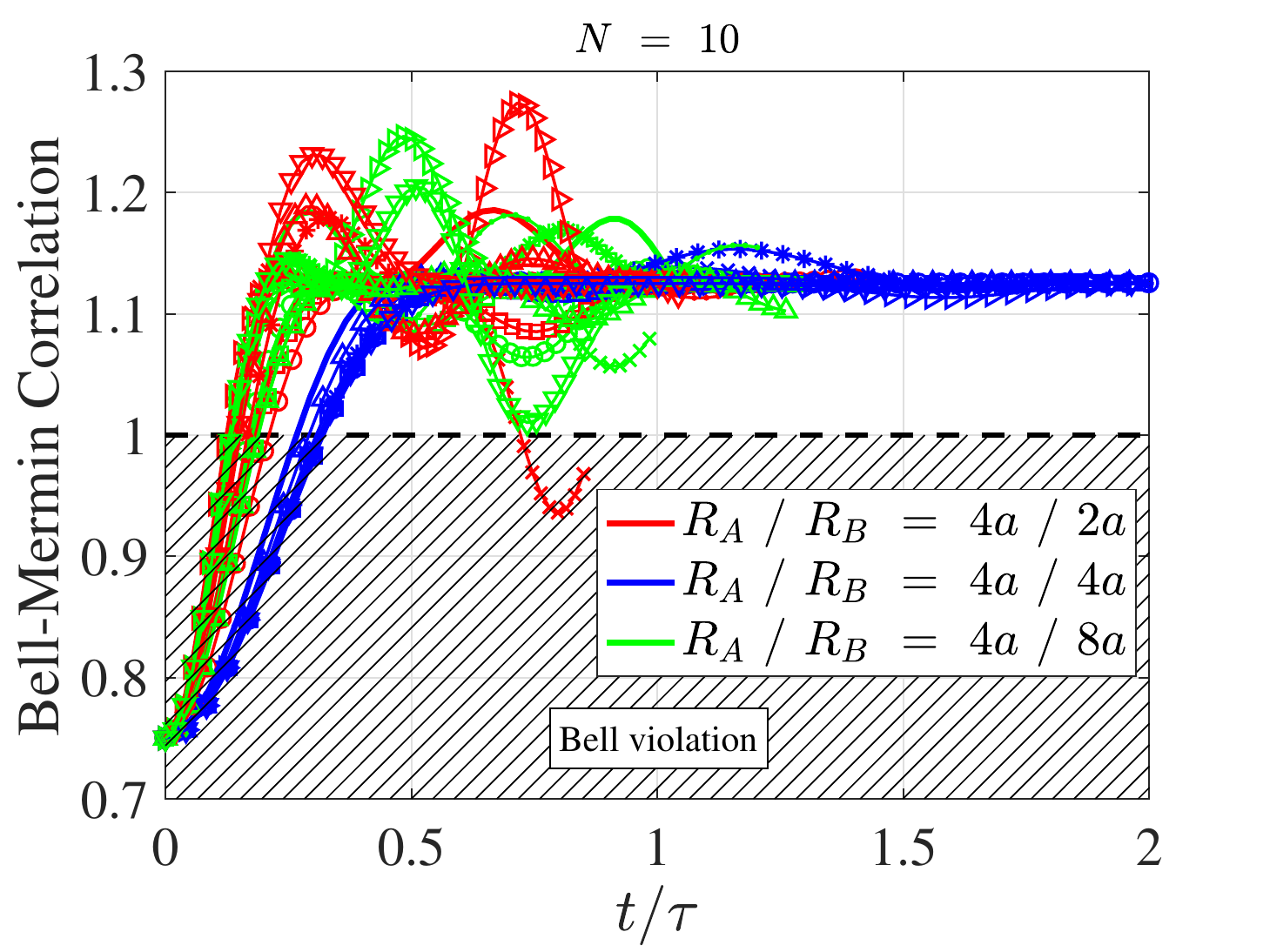}}
\par\end{centering}
\caption{A disentanglement time scale, $\tau$, characterizes the time scale
of disentanglement when the two local environments interact with their
individual target MCQ with roughly the same strength (that is, when
$R_{A}=R_{B}$); however, $\tau$ does not generalize to cases where
$R_{A}\protect\neq R_{B}$. Here, $a=1$ nm, and global environmental
population is $N=10$ for 3 different cases: $R_{A}/R_{B}\in\left\{ 4a/2a,4a/4a,4a/8a\right\} $.
(a) The BM correlation function, $S\TextSub{BM}$, is plotted against
time in fs for several time evolutions, and each randomized environment
drives a unique time evolution. (b) When $S\TextSub{BM}$ for each
evolution is plotted against time scaled to its own $\tau$, $\tau$
is only partially effective as a time scale. It is most effective
when $R_{A}=R_{B}$ (blue plots), mapping the various $R_{A}=R_{B}$
evolutions to roughly the same scaled time dependence. If $\tau$
also were an effective time scale for the $R_{A}\protect\neq R_{B}$
evolutions, the red and green plots would also overlay the blue plots.
However, $\tau$ overestimates the time constant when $R_{A}\protect\neq R_{B}$.\label{fig:differ_radius_BM_geo}}
\end{figure*}

The limitations of $\tau$ as a time scale for disentanglement are
illustrated in Figure \ref{fig:differ_radius_BM_geo}. Here, the local
environments are populated with $M=5$ DQDs each, and $S\TextSub{BM}$
for the target MCQ pair is plotted for several randomized environments
with different radial ratios, $R_{A}/R_{B}$. In particular, $R_{A}$
was fixed at $R_{A}=4a$ and $R_{B}\in\{R_{A}/2,R_{A},2R_{A}\}$ was
chosen with $a=1\;\text{nm}$.

In subplot \ref{fig:differ_radius_BM_geo}(a), $S\TextSub{BM}$ is
plotted versus time in fs for several random environments, and diverse
environmental interaction strengths drive disentanglement at diverse
speeds. A small $R_{B}$ results in strong $B$-$\mc{E}$ interactions
(red-line cases) and drives the fastest disentanglement, as $S\TextSub{BM}$
rapidly leaves the Bell violation region. On the other hand, a large
$R_{B}$ generally allows the target pair to retain entanglement longer
(green-line cases), up to the point where $R_{B}$ is so large that
environmental interactions are dominated by $A$-$\mc{E}$ interactions,
and changing $R_{B}$ no longer has a significant effect on overall
$AB$-$\mc{E}$ interactions.

When each time evolution from \ref{fig:differ_radius_BM_geo}(a) is
time-scaled to its own particular $\tau$, as in subfigure \ref{fig:differ_radius_BM_geo}(b),
the various time evolutions for the $R_{A}=R_{B}$ case roughly overlay
one another, having approximately the same time-scaled form (see the
blue plots). This is consistent with previous work,\citep{blair2018entanglement}
which suggests that $\tau$ is an effective time scale for characterizing
disentanglement when $R_{A}=R_{B}$. On the other hand, the $\tau$-scaled
calculations of $S\TextSub{BM}$ with $R_{A}\neq R_{B}$ do not overlay
the $\tau$-scaled $R_{A}=R_{B}$ plots, indicating that $\tau$ is
not as effective a time scale when $R_{A}\neq R_{B}$. For the $R_{A}\neq R_{B}$
cases, $\tau$ overestimates the time scale for disentanglement.

\begin{figure*}
\begin{centering}
\subfloat[]{\centering{}\includegraphics[width=0.48\textwidth,trim={0.75cm 0 1cm 0.65cm}, clip]{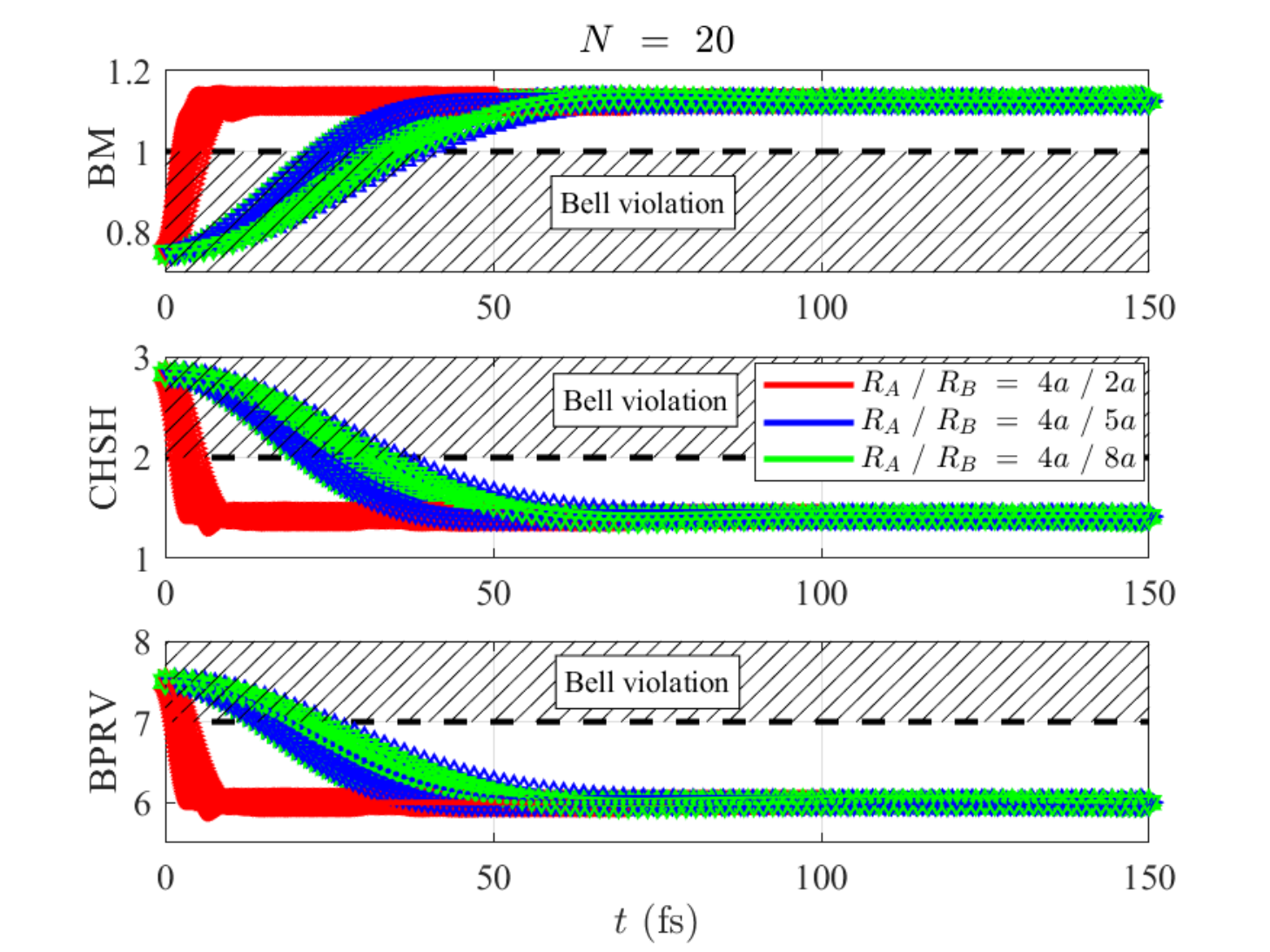}}
\hfill{}
\subfloat[]{\centering{}\includegraphics[width=0.48\textwidth,trim={0.75cm 0 1cm 0.65cm}, clip]{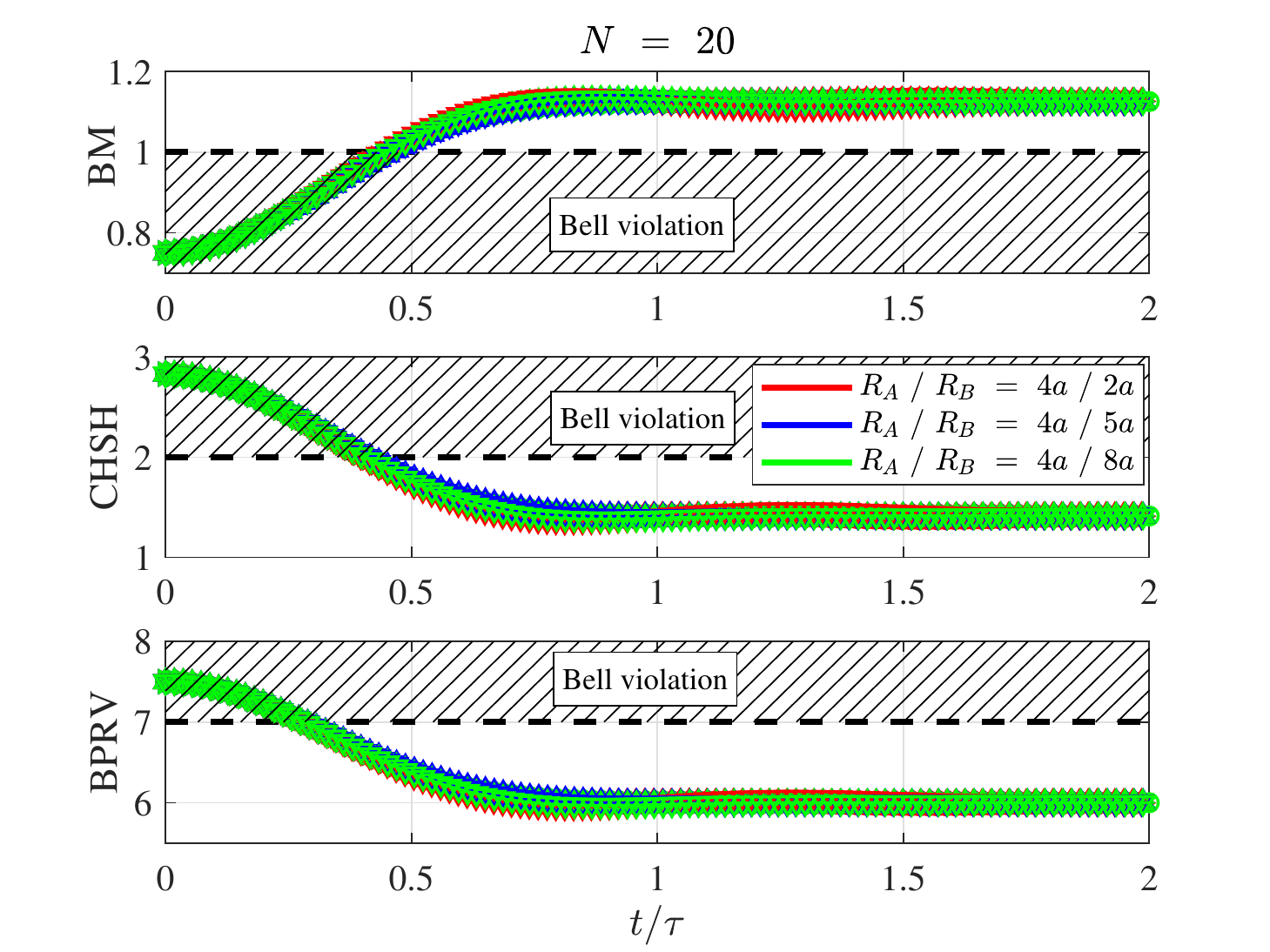}}
\par\end{centering}
\caption{The time scale $\tau_{E}$ of Equation (\ref{eq:TimeScaleTauE}) characterizes
the dynamics of environmentally-driven disentangement for various
ratios $R_{A}/R_{B}$. Here, $R_{A}$ is kept constant, and $R_{B}$
is varied with $a=1\,\text{nm}$ and $N=20$. (a) Correlation functions
from Section \ref{subsubsect:measures_of_ent} are plotted against
time in fs for several random environments, showing that varied $AB$-$\mc{E}$
interaction strengths drive disentanglement over varied durations.
(b) Each time evolution of subfigure (a) is time-scaled to its particular
$\tau_{E}$, resulting in a common time-scaled form within each correlation
function up to slight oscillations.\label{fig:BM_time_scale}}
\end{figure*}

Figure \ref{fig:BM_time_scale} shows $\tau_{E}$ of Equation (\ref{eq:TimeScaleTauE})
is effective at capturing the dynamics of disentanglement, even in
cases where $R_{A}\neq R_{B}$. Here, several time evolutions are
calculated, each for a different randomized environment. In each case,
we use $a=1$ nm, and $R_{A}=4a$, but $R_{B}$ is varied. For these
time evolutions, subfigure \ref{fig:BM_time_scale}(a) provides $S\TextSub{BM}$,
$S\TextSub{CHSH}$, and $S\TextSub{BPRV}$ plots against time in fs.
As expected, a diverse range of environmental interaction strengths
leads to diverse plots of the correlation functions with dynamics
on different time scales. When these plots are time-scaled to $\tau_{E}$,
as in subfigure \ref{fig:BM_time_scale}(b), the $\tau_{E}$-scaled
correlation function plots have a common form and overlay one another
for all $R_{A}/R_{B}$ ratios shown, neglecting long-time oscillations.
Indeed, $\tau_{E}$ characterizes well the dynamics of disentanglement.


\subsection{Early-time Gaussian Decay of Coherences}

\begin{figure*}
\begin{centering}
\subfloat[]{\centering{}\includegraphics[width=0.49\textwidth,trim={0.375cm 0 1.125cm 0},clip]{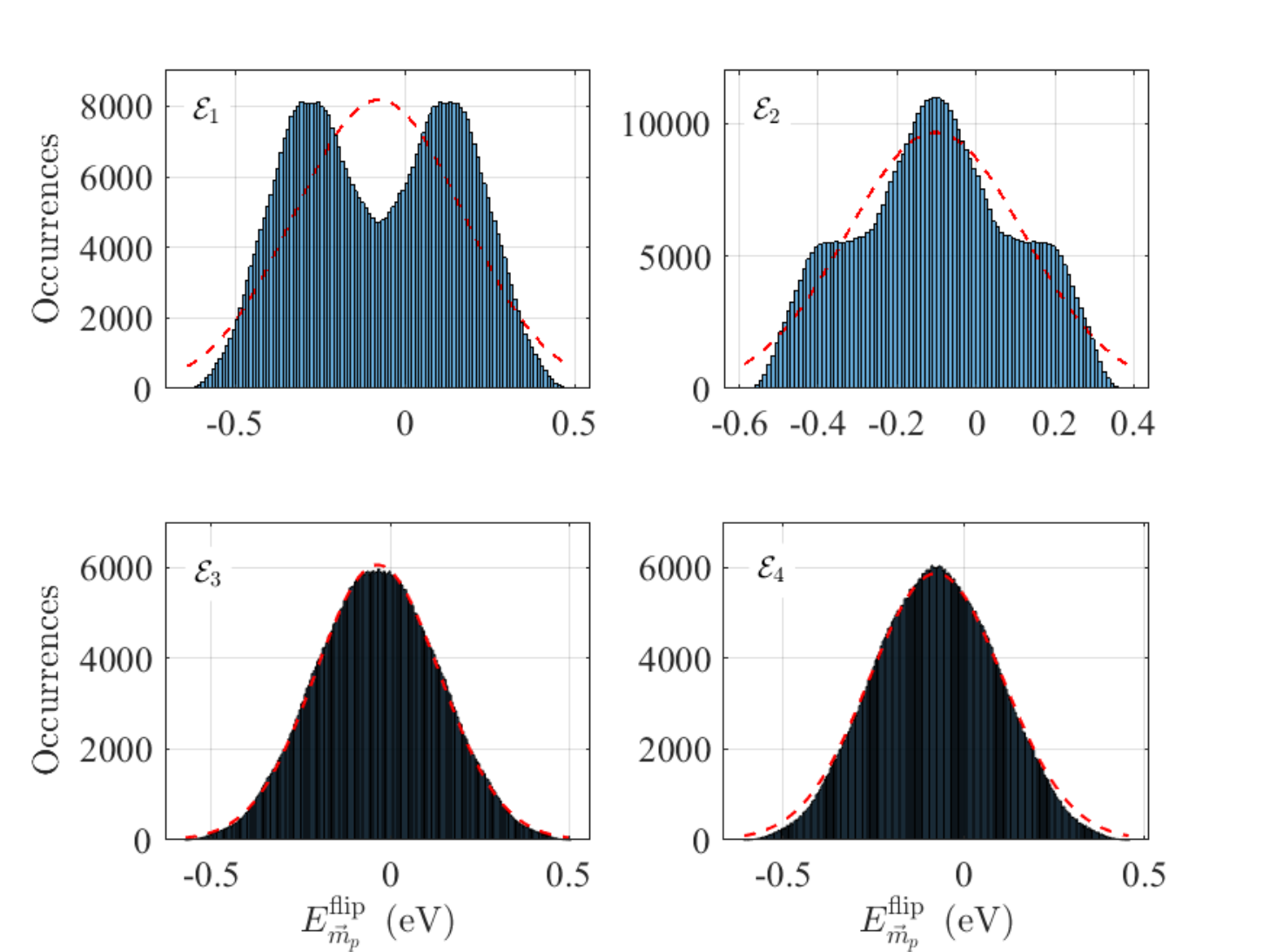}}
\hfill{}
\subfloat[]{\centering{}\includegraphics[width=0.49\textwidth,trim={0.375cm 0 1.125cmcm 0},clip]{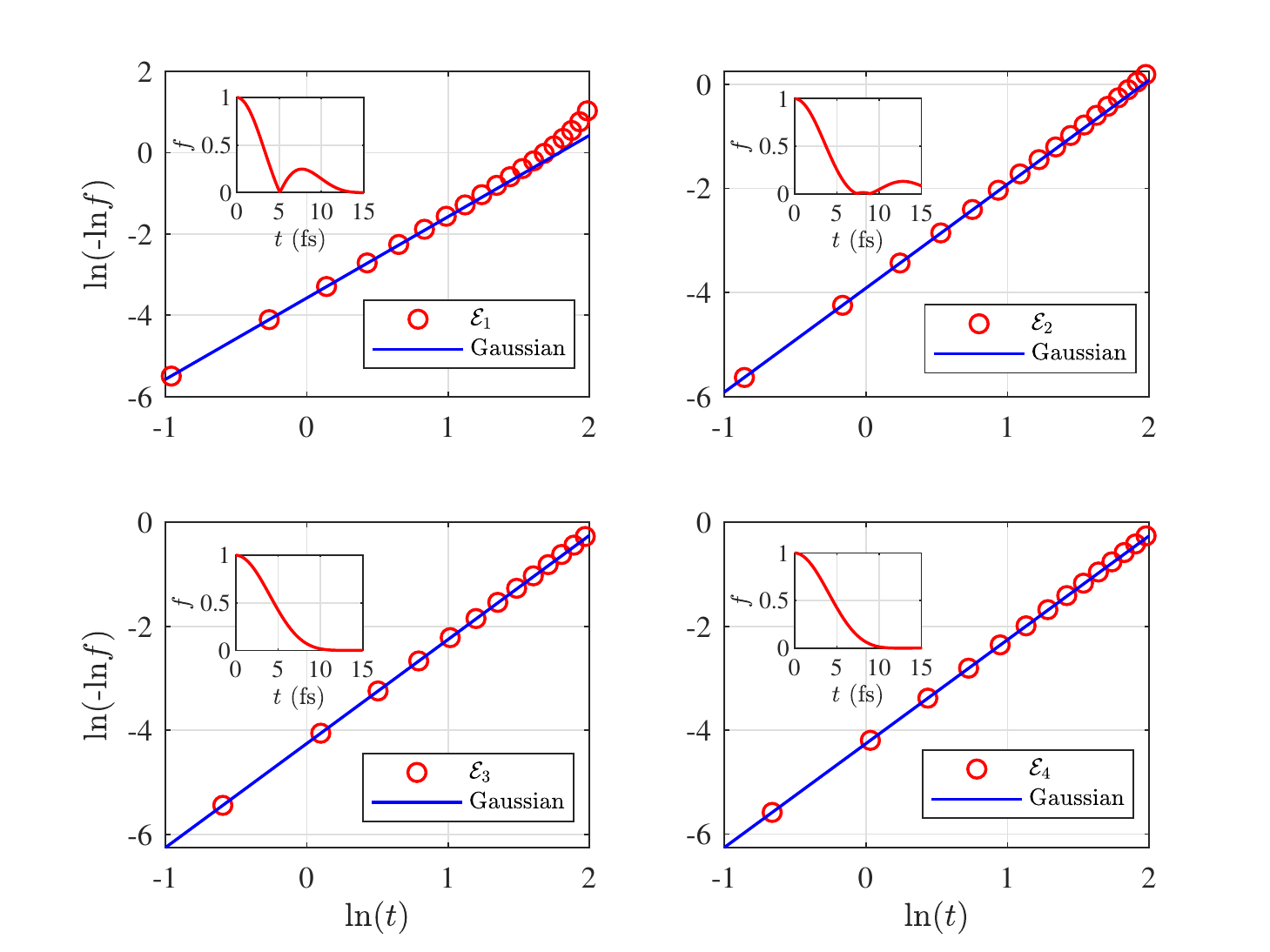}}
\par\end{centering}
\caption{Early-time behavior for both Gaussian and non-Gaussian $\{E\TextSup{flip}_{\vec{m}_{p}}\}$
distributions exhibit Gaussian decay in coherences at early times.
For each of the random environments with $\{E\TextSup{flip}_{\vec{m}_{p}}\}$
distributions plotted in subfigure (a), a linearization of the time-dependence
of he coherences is shown in subfigure (b). Data points that exhibit
Gaussian decay fall on the blue line of slope $dy/dx=2$, which marks
a truly Gaussian dependence. For all environments represented in (a),
early time points exhibit Gaussian decay. The time dependence of $f\left(t\right)$
is provided as an inset for each linearization panel. Here, environments
with $N=20$ neighbors were used, with $a=1\;\text{nm}$ and $R_{A}/R_{B}=4a/2a$.\label{fig:GaussianDecayOfCoherences}}
\end{figure*}

Figure \ref{fig:GaussianDecayOfCoherences} shows that the magnitudes
of the coherences $\Braket{00|\hat{\rho}_{AB}|11}=\Braket{11|\hat{\rho}_{AB}|00}^{\ast}$
generally exhibit a Gaussian decay in the early-time behavior, even
for $\{E\TextSup{flip}_{\vec{m}_{p}}\}$ distributions that deviate
from a zero-centric Gaussian distribution and cause notable revivals
in the magnitude of the coherences. To show this, a linearization
technique is applied to the $f\left(t\right)$ data. A Gaussian function
$g\left(t\right)=\exp\left(-t^{2}/2\sigma_{t}^{2}\right)$ may be
linearized to obtain
\[
\ln\left(-\ln g\right)=2\ln t-\ln\left(2\sigma_{t}^{2}\right).
\]
Therefore, a function $f\left(t\right)$ may be characterized as Gaussian
if a plot of $y=\ln\left(-\ln f\right)$ versus $x=\ln t$ has a slope
of $dy/dx=+2$. Four environments, $\{\mc{E}_{1},\mc{E}_{2},\mc{E}_{3},\mc{E}_{4}\}$
were selected and characterized. Their $\{E\TextSup{flip}_{\vec{m}_{p}}\}$
distributions are shown in subfigure \ref{fig:GaussianDecayOfCoherences}(a),
and the linearization of each $f\left(t\right)$ is plotted in \ref{fig:GaussianDecayOfCoherences}(b).
For each plot, a blue line of slope $+2$ (labeled ``Gaussian'')
is drawn through the left-most data point. Since several subsequent
linearized data points fall on or very close to the Gaussian marker
line, we say that these time evolutions are highly Gaussian, especially
at early times.

\section{Discussion\label{sec:Discussion}}

We discuss why the previously-used time scale, $\tau$, is suitable
when $R_{A}=R_{B}$ but becomes less suitable when $R_{A}\neq R_{B}$.

The time scale $\tau$ was defined as the geometric mean of time scales
$\tau_{A}$ and $\tau_{B}$,\citep{blair2018entanglement} which are
time scales for decoherence of a single MCQ, $A$ or $B$, in environments
$\mc{E}_{A}$ and $\mc{E}_{B}$, respectively:\citep{RamseyKrausOp2017}
\begin{equation}
\tau=\sqrt{\tau_{A}\tau_{B}}.\label{eq:TauGeometricMean}
\end{equation}
Each $\tau_{X}$ for $X\in\left\{ A,B\right\} $ was defined as
\begin{equation}
\tau_{X}=\frac{\pi\hbar}{E\TextSub{RMS}^{\left(X\right)}},\label{eq:TauX}
\end{equation}
where $E^{\left(X\right)}\TextSub{RMS}$ is the root-mean-square value
of the single-bit-flip energies $\{E_{X,j}\}$ in evironment $\mc{E}_{X}$
comprised of $M$ randomly-oriented DQDs randomly placed on the surface
of a shell of radius $R_{X}$ from the target MCQ:
\begin{equation}
E\TextSub{RMS}^{\left(X\right)}=\left(\frac{1}{2^{M}}\sum_{j=0}^{2^{M}-1}\left[E_{X}\left(\vec{m}_{X,j}\right)\right]^{2}\right)^{1/2}\,.\label{eq:EXRMS-00}
\end{equation}
Here, $E_{X}\left(\vec{m}_{X,j}\right)$ is the single-bit-flip energy
of the target MCQ given environmental state $\ket{\vec{m}_{X,j}}$,
labeled by the $M$-bit binary word
\[
\vec{m}_{X,j}=m_{M}m_{M-1}\cdots m_{2}m_{1}.
\]
Additionally, for each state $\ket{\vec{m}_{X,j}}$, there is a complementary
state $\ket{\vec{m}_{X,\bar{j}}}$,
\[
\vec{m}_{X,\bar{j}}=\bar{m}_{M}\bar{m}_{M-1}\cdots\bar{m}_{k}\cdots\bar{m}_{2}\bar{m}_{1}\;,
\]
for which the label $\vec{m}_{X,\bar{j}}$ is the bit-wise complement
of $\vec{m}_{X,j}$, and for which $E_{X}\left(\vec{m}_{X,\bar{j}}\right)=-E_{X}\left(\vec{m}_{X,j}\right)$.
Now, let us order $\{E_{X}\left(\vec{m}_{X,\bar{j}}\right)\}$ from
most positive to most negative, and then relabel this ordered set
$\{\varepsilon_{X,a}\}$, where $a$ is a counting number smaller
than $2^{M}$. It is now possible to write $E\TextSub{RMS}^{\left(X\right)}$
of Equation (\ref{eq:EXRMS-00}) in terms of only the first $2^{M-1}$
energies $\{\varepsilon_{X,a}\}$, which are non-negative by virtue
of ordering:
\begin{equation}
E\TextSub{RMS}^{\left(X\right)}=\frac{1}{2^{(M-1)/2}}\left(\sum_{a=0}^{2^{M-1}-1}\varepsilon_{X,a}^{2}\right)^{1/2}.\label{eq:EXRMS}
\end{equation}
Then, $\tau=\sqrt{\tau_{A}\tau_{B}}$ is found by combining Equations
(\ref{eq:TauGeometricMean}), (\ref{eq:TauX}), and (\ref{eq:EXRMS}):
\begin{equation}
\tau=\frac{\pi\hbar2^{(M-1)/2}}{\left(\left(\sum_{a=0}^{2^{M-1}-1}\varepsilon_{A,a}^{2}\right)^{1/2}\left(\sum_{b=0}^{2^{M-1}-1}\varepsilon_{B,b}^{2}\right)^{1/2}\right)^{1/2}}\;.\label{eq:TauGeometricExplicit}
\end{equation}

On the other hand, there are $2^{2M}$ double-bit-flip energies $\{E_{\vec{m}_{p}}^{\text{flip}}\}$
as defined in Equation (\ref{eq:DoubleBitFlipEnergy}). These double-bit-flip
energies can be formed by adding and subtracting only the positive
single bit flip energies $\pm\varepsilon_{A,a}$ to $\pm\varepsilon_{B,b}$,
since $A$ and $B$ do not interact: $\{E_{\vec{m}_{p}}^{\text{flip}}\}=\{\pm\varepsilon_{A,a}\pm\varepsilon_{B,b}\}$.
It can be shown that the RMS value of these double-bit-flip energies
is given by
\begin{equation}
E\TextSup{flip}\TextSub{RMS}=\frac{1}{2^{(M-1)/2}}\left(\sum_{j=0}^{2^{M-1}-1}\varepsilon_{A,j}^{2}+\varepsilon_{B,j}^{2}\right)^{1/2}\;.\label{eq:EflipRMS_EaEb}
\end{equation}
Thus, by combining Equations (\ref{eq:TimeScaleTauE}) and (\ref{eq:EFlipRMS}),
the time scale $\tau_{E}$ may be written as:
\begin{equation}
\tau_{E}=\frac{\pi\hbar2^{(M-1)/2}}{\left(\sum_{j=0}^{2^{M-1}-1}\varepsilon_{A,j}^{2}+\varepsilon_{B,j}^{2}\right)^{1/2}}\label{eq:TimeScaleTauE_v02}
\end{equation}
Here, $\tau_{E}$ is written without any cross-terms, i.e. without
products $\varepsilon_{A,a}^{m}\varepsilon_{B,b}^{n}$. That the total
energies of interest in $\tau_{E}$ are sums of the non-negative energies
$\varepsilon_{A,a}$ and $\varepsilon_{B,b}$ (and their powers) reflects
the fact that $A$ and $B$ do not interact. On the other hand, cross-terms
arise in the $\tau$ of Equation (\ref{eq:TauGeometricExplicit}).
Only when $\varepsilon_{B,j}\rightarrow\varepsilon_{A,j}$ do the
cross-terms vanish from $\tau$. This is achieved approximately in
our global system when $R_{A}=R_{B}$. In this case we can take the
ratio of Equations (\ref{eq:TimeScaleTauE_v02}) and (\ref{eq:TauGeometricExplicit})
is $\tau_{E}/\tau\rightarrow1/\sqrt{2}$, and $\tau$ becomes approximately
proportional to $\tau_{E}$. This proportionality between $\tau$
and $\tau_{E}$ allows $\tau$ to function as a characteristic time
constant for the dynamics of disentanglement in the $R_{A}=R_{B}$
limit, as seen in previous work.\citep{blair2018entanglement}

\begin{figure}
\begin{centering}
\includegraphics[width=1\columnwidth]{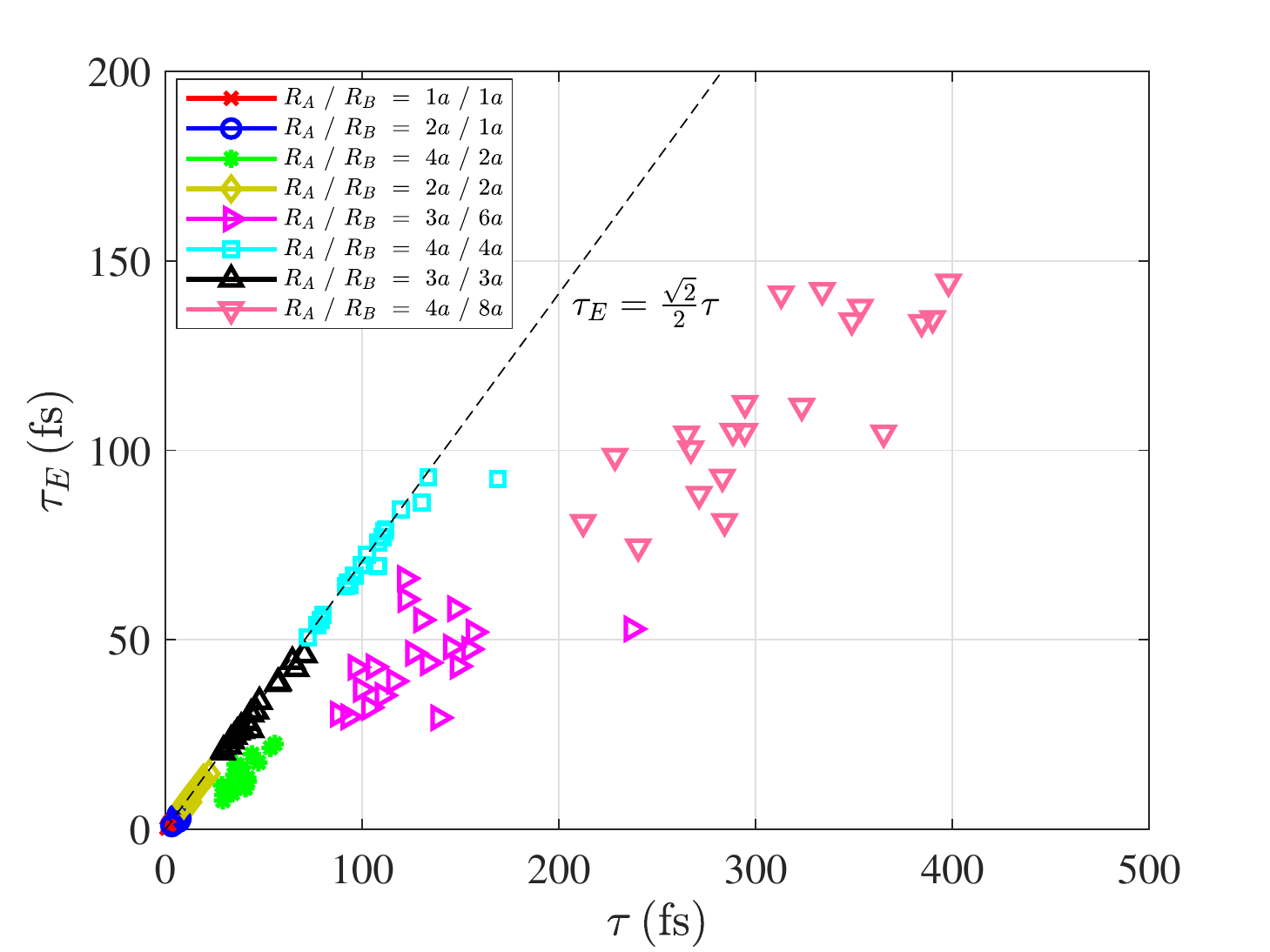}
\par\end{centering}
\caption{When $R_{A}=R_{B}$, $\tau\rightarrow\tau_{E}\sqrt{2}$, and $\tau$
functions as an effective time constant for the loss of disentanglement
because $\tau\propto\tau_{E}$. When $R_{A}\protect\neq R_{B}$, $\tau$
deviates from proporionality to $\tau_{E}$, becoming less effective
as a time constant. Here, $\tau_{E}$ and $\tau$ were calculated
for several randomized environments for various $R_{A}/R_{B}$ ratios,
each with $a=1~\mbox{nm}$ and an environmental population $N=10$.\label{fig:scatter_tauE_v_tau}}
\end{figure}

On the other hand, when $R_{A}\neq R_{B}$, the proportionality between
$\tau_{E}$ and $\tau$ is lost, and $\tau$ fails as a characteristic
time constant. This relationship is shown in the data of Figure \ref{fig:scatter_tauE_v_tau}.
Here, a scatter plot is made for $\tau_{E}$ and $\tau$ data for
various ratios of $R_{A}/R_{B}$ and several randomized environments
for each ratio. When $R_{A}=R_{B}$, the points of the scatter plot
fall close to the line $\tau_{E}=\tau/\sqrt{2}$; but, when $R_{A}\neq R_{B}$
the data departs from that proportionality. Mathematically, this is
driven by the unphysical cross-terms arising in the approximate time
constant $\tau$ when $R_{A}\neq R_{B}$.

\section{Conclusion\label{sec:Conclusion}}

In this paper, the dynamics of the loss of entanglement are studied
in MCQs. Each qubit is immersed in its own local environment, modeled
using a set of $M$ neighboring DQDs. Tunneling also is suppressed
to eliminate dissipative effects and leave only entanglement. The
system-environment interactions drive the the gradual loss of entanglement
between $A$ and $B$. The loss of entanglement has a Gaussian from,
especially at early times. This behavior is not reproducible using
Markovian models of memoryless environments, which can yield only
an exponential time-dependence; however, we have developed non-Markovian
models, including an exact semi-analytic model and an approximate
Gaussian model for the density operator of the target MCQ pair. The
time scale of this disentanglement is directly related to the the
strength of the electrostatic interaction between the environment
and the target pair of qubits. This time scale, $\tau_{E}$, describes
this problem in a more general way than does a previously-developed
time scale, $\tau$. While $\tau$ is useful in the case where each
local environment an approximately equal strength of interaction its
central target MCQ (the local environments have the same radius),
$\tau_{E}$ also characterizes systems where environmental interactions
are dominant for only one MCQ in the target pair (the weaker environment
has a larger radius than the dominant local environment). The approximate
Gaussian model may be used to provide an accurate, non-Markovian description
of system dynamics under the influence of a much broader class of
environments characterized by randomness. Models of disentanglement
and other quantum phenomena can help explore the dynamics of MCQs
and the role they can play in quantum information processing under
the influence of the environment.

\begin{acknowledgments}
The authors thank Craig S. Lent from the University of Notre Dame
for engaging with us in dialogue on this work. We also gratefully
acknowledge peer reviewers for insightful ideas and comments, which
were helpful in developing the models presented here. This work was
sponsored by Baylor University under a new-faculty startup grant.
\end{acknowledgments}


\bibliography{MCQ_Disentanglement}

\providecommand{\noopsort}[1]{}\providecommand{\singleletter}[1]{#1}%
\begin{thebibliography}{22}%
\makeatletter
\providecommand \@ifxundefined [1]{%
 \@ifx{#1\undefined}
}%
\providecommand \@ifnum [1]{%
 \ifnum #1\expandafter \@firstoftwo
 \else \expandafter \@secondoftwo
 \fi
}%
\providecommand \@ifx [1]{%
 \ifx #1\expandafter \@firstoftwo
 \else \expandafter \@secondoftwo
 \fi
}%
\providecommand \natexlab [1]{#1}%
\providecommand \enquote  [1]{``#1''}%
\providecommand \bibnamefont  [1]{#1}%
\providecommand \bibfnamefont [1]{#1}%
\providecommand \citenamefont [1]{#1}%
\providecommand \href@noop [0]{\@secondoftwo}%
\providecommand \href [0]{\begingroup \@sanitize@url \@href}%
\providecommand \@href[1]{\@@startlink{#1}\@@href}%
\providecommand \@@href[1]{\endgroup#1\@@endlink}%
\providecommand \@sanitize@url [0]{\catcode `\\12\catcode `\$12\catcode
  `\&12\catcode `\#12\catcode `\^12\catcode `\_12\catcode `\%12\relax}%
\providecommand \@@startlink[1]{}%
\providecommand \@@endlink[0]{}%
\providecommand \url  [0]{\begingroup\@sanitize@url \@url }%
\providecommand \@url [1]{\endgroup\@href {#1}{\urlprefix }}%
\providecommand \urlprefix  [0]{URL }%
\providecommand \Eprint [0]{\href }%
\providecommand \doibase [0]{http://dx.doi.org/}%
\providecommand \selectlanguage [0]{\@gobble}%
\providecommand \bibinfo  [0]{\@secondoftwo}%
\providecommand \bibfield  [0]{\@secondoftwo}%
\providecommand \translation [1]{[#1]}%
\providecommand \BibitemOpen [0]{}%
\providecommand \bibitemStop [0]{}%
\providecommand \bibitemNoStop [0]{.\EOS\space}%
\providecommand \EOS [0]{\spacefactor3000\relax}%
\providecommand \BibitemShut  [1]{\csname bibitem#1\endcsname}%
\let\auto@bib@innerbib\@empty
\bibitem [{\citenamefont {Feynman}(1982)}]{FeynmanQuantumSimulation1982}%
  \BibitemOpen
  \bibfield  {author} {\bibinfo {author} {\bibfnamefont {R.}~\bibnamefont
  {Feynman}},\ }\href@noop {} {\bibfield  {journal} {\bibinfo  {journal} {Int.\
  J.\ Theor.\ Phys.}\ }\textbf {\bibinfo {volume} {21}},\ \bibinfo {pages}
  {467} (\bibinfo {year} {1982})}\BibitemShut {NoStop}%
\bibitem [{\citenamefont {Feynman}(1985)}]{FeynmanGenesis1985}%
  \BibitemOpen
  \bibfield  {author} {\bibinfo {author} {\bibfnamefont {R.}~\bibnamefont
  {Feynman}},\ }\href@noop {} {\bibfield  {journal} {\bibinfo  {journal}
  {Optics news}\ }\textbf {\bibinfo {volume} {11}},\ \bibinfo {pages} {11}
  (\bibinfo {year} {1985})}\BibitemShut {NoStop}%
\bibitem [{\citenamefont {Shor}(1994)}]{shor1994algorithms}%
  \BibitemOpen
  \bibfield  {author} {\bibinfo {author} {\bibfnamefont {P.~W.}\ \bibnamefont
  {Shor}},\ }in\ \href@noop {} {\emph {\bibinfo {booktitle} {Foundations of
  Computer Science, 1994 Proceedings., 35th Annual Symposium on}}}\ (\bibinfo
  {organization} {Ieee},\ \bibinfo {year} {1994})\ pp.\ \bibinfo {pages}
  {124--134}\BibitemShut {NoStop}%
\bibitem [{\citenamefont {Grover}(1996)}]{GroverSearchUnpub}%
  \BibitemOpen
  \bibfield  {author} {\bibinfo {author} {\bibfnamefont {L.}~\bibnamefont
  {Grover}},\ }\href {\doibase arXiv:quant-ph/9605043v3} {\bibfield  {journal}
  {\bibinfo  {journal} {unpublished}\ } (\bibinfo {year} {1996}),\
  arXiv:quant-ph/9605043v3}\BibitemShut {NoStop}%
\bibitem [{\citenamefont {Farhi}\ \emph {et~al.}(2000)\citenamefont {Farhi},
  \citenamefont {Goldstone}, \citenamefont {Gutmann},\ and\ \citenamefont
  {Sipser}}]{adiabaticQC}%
  \BibitemOpen
  \bibfield  {author} {\bibinfo {author} {\bibfnamefont {E.}~\bibnamefont
  {Farhi}}, \bibinfo {author} {\bibfnamefont {J.}~\bibnamefont {Goldstone}},
  \bibinfo {author} {\bibfnamefont {S.}~\bibnamefont {Gutmann}}, \ and\
  \bibinfo {author} {\bibfnamefont {M.}~\bibnamefont {Sipser}},\ }\href
  {\doibase arXiv:quant-ph/0001106v1} {\bibfield  {journal} {\bibinfo
  {journal} {unpublished}\ } (\bibinfo {year} {2000}),\
  arXiv:quant-ph/0001106v1}\BibitemShut {NoStop}%
\bibitem [{\citenamefont {Bennet}\ and\ \citenamefont {Brassard}(1984)}]{BB84}%
  \BibitemOpen
  \bibfield  {author} {\bibinfo {author} {\bibfnamefont {C.}~\bibnamefont
  {Bennet}}\ and\ \bibinfo {author} {\bibfnamefont {G.}~\bibnamefont
  {Brassard}},\ }in\ \href@noop {} {\emph {\bibinfo {booktitle} {Proceedings of
  the IEEE International Conference on Computers, Systems and Signal
  Processing}}}\ (\bibinfo {year} {1984})\ pp.\ \bibinfo {pages}
  {175--179}\BibitemShut {NoStop}%
\bibitem [{\citenamefont {Ekert}(1991)}]{QuantumCryptography1991}%
  \BibitemOpen
  \bibfield  {author} {\bibinfo {author} {\bibfnamefont {A.}~\bibnamefont
  {Ekert}},\ }\href@noop {} {\bibfield  {journal} {\bibinfo  {journal} {Phys
  Rev Lett}\ }\textbf {\bibinfo {volume} {67}},\ \bibinfo {pages} {661}
  (\bibinfo {year} {1991})}\BibitemShut {NoStop}%
\bibitem [{\citenamefont {Jozsa}\ and\ \citenamefont
  {Linden}(2003)}]{jozsa2003role}%
  \BibitemOpen
  \bibfield  {author} {\bibinfo {author} {\bibfnamefont {R.}~\bibnamefont
  {Jozsa}}\ and\ \bibinfo {author} {\bibfnamefont {N.}~\bibnamefont {Linden}},\
  }\href@noop {} {\bibfield  {journal} {\bibinfo  {journal} {Proceedings of the
  Royal Society of London. Series A: Mathematical, Physical and Engineering
  Sciences}\ }\textbf {\bibinfo {volume} {459}},\ \bibinfo {pages} {2011}
  (\bibinfo {year} {2003})}\BibitemShut {NoStop}%
\bibitem [{\citenamefont {Mujica-Martinez}, \citenamefont {Nalbach},\ and\
  \citenamefont {Thorwart}(2013)}]{2013_MCQ}%
  \BibitemOpen
  \bibfield  {author} {\bibinfo {author} {\bibfnamefont {C.}~\bibnamefont
  {Mujica-Martinez}}, \bibinfo {author} {\bibfnamefont {P.}~\bibnamefont
  {Nalbach}}, \ and\ \bibinfo {author} {\bibfnamefont {M.}~\bibnamefont
  {Thorwart}},\ }\href {\doibase doi:10.1103/PhysRevLett.111.016802} {\bibfield
   {journal} {\bibinfo  {journal} {Phys.\ Rev.\ Lett.}\ }\textbf {\bibinfo
  {volume} {111}},\ \bibinfo {pages} {016802} (\bibinfo {year}
  {2013})}\BibitemShut {NoStop}%
\bibitem [{\citenamefont {Lent}(2000)}]{LentScience2000}%
  \BibitemOpen
  \bibfield  {author} {\bibinfo {author} {\bibfnamefont {C.~S.}\ \bibnamefont
  {Lent}},\ }\href@noop {} {\bibfield  {journal} {\bibinfo  {journal}
  {Science}\ }\textbf {\bibinfo {volume} {288}},\ \bibinfo {pages} {1597}
  (\bibinfo {year} {2000})}\BibitemShut {NoStop}%
\bibitem [{\citenamefont {T{\'o}th}\ and\ \citenamefont
  {Lent}(2001)}]{QCA_QuantComp_2001}%
  \BibitemOpen
  \bibfield  {author} {\bibinfo {author} {\bibfnamefont {G.}~\bibnamefont
  {T{\'o}th}}\ and\ \bibinfo {author} {\bibfnamefont {C.}~\bibnamefont
  {Lent}},\ }\href {\doibase 10.1103/PhysRevA.63.052315} {\bibfield  {journal}
  {\bibinfo  {journal} {Phys. Rev. A}\ }\textbf {\bibinfo {volume} {63}}
  (\bibinfo {year} {2001}),\ 10.1103/PhysRevA.63.052315}\BibitemShut {NoStop}%
\bibitem [{\citenamefont {Lieberman}\ \emph {et~al.}(2002)\citenamefont
  {Lieberman}, \citenamefont {Chellamma}, \citenamefont {Varughese},
  \citenamefont {Wang}, \citenamefont {Lent}, \citenamefont {Bernstein},
  \citenamefont {Snider},\ and\ \citenamefont
  {Peiris}}]{QCA_at_molecular_scale}%
  \BibitemOpen
  \bibfield  {author} {\bibinfo {author} {\bibfnamefont {M.}~\bibnamefont
  {Lieberman}}, \bibinfo {author} {\bibfnamefont {S.}~\bibnamefont
  {Chellamma}}, \bibinfo {author} {\bibfnamefont {B.}~\bibnamefont
  {Varughese}}, \bibinfo {author} {\bibfnamefont {Y.}~\bibnamefont {Wang}},
  \bibinfo {author} {\bibfnamefont {C.}~\bibnamefont {Lent}}, \bibinfo {author}
  {\bibfnamefont {G.}~\bibnamefont {Bernstein}}, \bibinfo {author}
  {\bibfnamefont {G.}~\bibnamefont {Snider}}, \ and\ \bibinfo {author}
  {\bibfnamefont {F.}~\bibnamefont {Peiris}},\ }\href@noop {} {\bibfield
  {journal} {\bibinfo  {journal} {Ann. N.Y. Acad. Sci.}\ }\textbf {\bibinfo
  {volume} {960}},\ \bibinfo {pages} {225} (\bibinfo {year}
  {2002})}\BibitemShut {NoStop}%
\bibitem [{\citenamefont {Blair}, \citenamefont {T{\'o}th},\ and\ \citenamefont
  {Lent}(2018)}]{blair2018entanglement}%
  \BibitemOpen
  \bibfield  {author} {\bibinfo {author} {\bibfnamefont {E.~P.}\ \bibnamefont
  {Blair}}, \bibinfo {author} {\bibfnamefont {G.}~\bibnamefont {T{\'o}th}}, \
  and\ \bibinfo {author} {\bibfnamefont {C.~S.}\ \bibnamefont {Lent}},\
  }\href@noop {} {\bibfield  {journal} {\bibinfo  {journal} {Journal of
  Physics: Condensed Matter}\ }\textbf {\bibinfo {volume} {30}},\ \bibinfo
  {pages} {195602} (\bibinfo {year} {2018})}\BibitemShut {NoStop}%
\bibitem [{\citenamefont {Lu}\ and\ \citenamefont
  {Lent}(2013)}]{2013_zwitterionicQCA_DQD}%
  \BibitemOpen
  \bibfield  {author} {\bibinfo {author} {\bibfnamefont {Y.}~\bibnamefont
  {Lu}}\ and\ \bibinfo {author} {\bibfnamefont {C.}~\bibnamefont {Lent}},\
  }\href {\doibase 10.1016/j.cplett.2013.07.019} {\bibfield  {journal}
  {\bibinfo  {journal} {Chem. Phys. Lett.}\ }\textbf {\bibinfo {volume}
  {582}},\ \bibinfo {pages} {86} (\bibinfo {year} {2013})}\BibitemShut
  {NoStop}%
\bibitem [{\citenamefont {Blair}, \citenamefont {Corcelli},\ and\ \citenamefont
  {Lent}(2016)}]{molecularQCAelectronTransfer}%
  \BibitemOpen
  \bibfield  {author} {\bibinfo {author} {\bibfnamefont {E.}~\bibnamefont
  {Blair}}, \bibinfo {author} {\bibfnamefont {S.}~\bibnamefont {Corcelli}}, \
  and\ \bibinfo {author} {\bibfnamefont {C.}~\bibnamefont {Lent}},\ }\href@noop
  {} {\bibfield  {journal} {\bibinfo  {journal} {J.\ Chem.\ Phys.}\ }\textbf
  {\bibinfo {volume} {145}},\ \bibinfo {pages} {014307} (\bibinfo {year}
  {2016})}\BibitemShut {NoStop}%
\bibitem [{\citenamefont {Lu}\ and\ \citenamefont
  {Lent}(2011)}]{2011_zwitterions}%
  \BibitemOpen
  \bibfield  {author} {\bibinfo {author} {\bibfnamefont {Y.}~\bibnamefont
  {Lu}}\ and\ \bibinfo {author} {\bibfnamefont {C.}~\bibnamefont {Lent}},\
  }\href@noop {} {\bibfield  {journal} {\bibinfo  {journal} {Phys. Chem. Chem.
  Phys.}\ }\textbf {\bibinfo {volume} {13}},\ \bibinfo {pages} {14928}
  (\bibinfo {year} {2011})}\BibitemShut {NoStop}%
\bibitem [{\citenamefont {Christie}\ \emph {et~al.}(2015)\citenamefont
  {Christie}, \citenamefont {Forrest}, \citenamefont {Corcelli}, \citenamefont
  {Wasio}, \citenamefont {Quardokus}, \citenamefont {Brown}, \citenamefont
  {Kandel}, \citenamefont {Lu}, \citenamefont {Lent},\ and\ \citenamefont
  {Henderson}}]{Christie15}%
  \BibitemOpen
  \bibfield  {author} {\bibinfo {author} {\bibfnamefont {J.}~\bibnamefont
  {Christie}}, \bibinfo {author} {\bibfnamefont {R.}~\bibnamefont {Forrest}},
  \bibinfo {author} {\bibfnamefont {S.}~\bibnamefont {Corcelli}}, \bibinfo
  {author} {\bibfnamefont {N.}~\bibnamefont {Wasio}}, \bibinfo {author}
  {\bibfnamefont {R.}~\bibnamefont {Quardokus}}, \bibinfo {author}
  {\bibfnamefont {R.}~\bibnamefont {Brown}}, \bibinfo {author} {\bibfnamefont
  {S.}~\bibnamefont {Kandel}}, \bibinfo {author} {\bibfnamefont
  {Y.}~\bibnamefont {Lu}}, \bibinfo {author} {\bibfnamefont {C.}~\bibnamefont
  {Lent}}, \ and\ \bibinfo {author} {\bibfnamefont {K.}~\bibnamefont
  {Henderson}},\ }\href@noop {} {\bibfield  {journal} {\bibinfo  {journal}
  {Angew. Chem. Int. Ed.}\ }\textbf {\bibinfo {volume} {54}},\ \bibinfo {pages}
  {15448} (\bibinfo {year} {2015})}\BibitemShut {NoStop}%
\bibitem [{Note1()}]{Note1}%
  \BibitemOpen
  \bibinfo {note} {The MCQs and the environmental molecules all are assumed to
  be DQDs of the same molecular species. However, for clarity, ``MCQ'' is
  reserved for the target pair of DQDs used to model qubits; on the other hand,
  ``DQD'' is more general and may be applied to both target molecules and
  environmental molecules. Following this train of thought, we reserve the term
  ``computational basis'' to describe fully-localized electronic states of the
  MCQs in $AB$, but the term ``classical basis'' could describe an analogous
  state in any system of DQDs\textendash either MCQ or
  environmental.}\BibitemShut {Stop}%
\bibitem [{\citenamefont {Mermin}(1985)}]{Mermin1985}%
  \BibitemOpen
  \bibfield  {author} {\bibinfo {author} {\bibfnamefont {N.~D.}\ \bibnamefont
  {Mermin}},\ }\href@noop {} {\bibfield  {journal} {\bibinfo  {journal}
  {Physics today}\ }\textbf {\bibinfo {volume} {4}},\ \bibinfo {pages} {38}
  (\bibinfo {year} {1985})}\BibitemShut {NoStop}%
\bibitem [{\citenamefont {Clauser}\ \emph {et~al.}(1969)\citenamefont
  {Clauser}, \citenamefont {Horne}, \citenamefont {Shimony},\ and\
  \citenamefont {Holt}}]{clauser1969proposed}%
  \BibitemOpen
  \bibfield  {author} {\bibinfo {author} {\bibfnamefont {J.~F.}\ \bibnamefont
  {Clauser}}, \bibinfo {author} {\bibfnamefont {M.~A.}\ \bibnamefont {Horne}},
  \bibinfo {author} {\bibfnamefont {A.}~\bibnamefont {Shimony}}, \ and\
  \bibinfo {author} {\bibfnamefont {R.~A.}\ \bibnamefont {Holt}},\ }\href@noop
  {} {\bibfield  {journal} {\bibinfo  {journal} {Physical review letters}\
  }\textbf {\bibinfo {volume} {23}},\ \bibinfo {pages} {880} (\bibinfo {year}
  {1969})}\BibitemShut {NoStop}%
\bibitem [{\citenamefont {\v{C}. Brukner}\ \emph {et~al.}(2006)\citenamefont
  {\v{C}. Brukner}, \citenamefont {Paunkovi\'{c}}, \citenamefont {Rudolph},\
  and\ \citenamefont {Vedral}}]{brukner2006entanglement}%
  \BibitemOpen
  \bibfield  {author} {\bibinfo {author} {\bibnamefont {\v{C}. Brukner}},
  \bibinfo {author} {\bibfnamefont {N.}~\bibnamefont {Paunkovi\'{c}}}, \bibinfo
  {author} {\bibfnamefont {T.}~\bibnamefont {Rudolph}}, \ and\ \bibinfo
  {author} {\bibfnamefont {V.}~\bibnamefont {Vedral}},\ }\href@noop {}
  {\bibfield  {journal} {\bibinfo  {journal} {International Journal of Quantum
  Information}\ }\textbf {\bibinfo {volume} {4}},\ \bibinfo {pages} {365}
  (\bibinfo {year} {2006})}\BibitemShut {NoStop}%
\bibitem [{\citenamefont {Ramsey}\ and\ \citenamefont
  {Blair}(2017)}]{RamseyKrausOp2017}%
  \BibitemOpen
  \bibfield  {author} {\bibinfo {author} {\bibfnamefont {J.}~\bibnamefont
  {Ramsey}}\ and\ \bibinfo {author} {\bibfnamefont {E.}~\bibnamefont {Blair}},\
  }\href@noop {} {\bibfield  {journal} {\bibinfo  {journal} {J. Appl. Phys.}\
  }\textbf {\bibinfo {volume} {122}},\ \bibinfo {pages} {084304} (\bibinfo
  {year} {2017})}\BibitemShut {NoStop}%
\end{thebibliography}%

\end{document}